\PassOptionsToPackage{dvipsnames}{xcolor}
 \documentclass[letterpaper,11pt]{article}
\usepackage{amsmath, amsthm, amssymb}
\usepackage[usenames, dvipsnames]{color}
\usepackage[normalem]{ulem} 
\usepackage{fullpage}
\usepackage{cite}
\usepackage[numbers, sort&compress]{natbib}
\usepackage[noend]{algorithmic}
\usepackage{tikz, pgfplots}
\usepackage{enumerate}
\usepackage{hyperref}
\usepackage{xspace,color}
\usepackage{graphicx}
\usepackage{caption}
\usepackage{subcaption}
\usepackage{enumitem,linegoal}
\usepackage[linesnumbered,ruled,vlined]{algorithm2e}
\usepackage{comment}
\usepackage{ctable}

\usepackage{mathtools}
\usepackage[flushleft]{threeparttable}
\usepackage{verbatim}
\usepackage{setspace}
\usepackage{bbm}
\usepackage{thm-restate}
\usepackage{footnote}
\usepackage[nameinlink,capitalize,noabbrev]{cleveref}
\crefname{LP}{LP}{LPs}
\usepackage{tikz}
\usetikzlibrary{positioning, decorations.pathmorphing}
\usepackage{pgfplots}
\pgfplotsset{compat=1.18}
\usepackage{hyperref}
\usepackage{tikz}
\usetikzlibrary{decorations.pathreplacing}
\usetikzlibrary{arrows.meta,fit}
\usepackage{booktabs,array}

\makesavenoteenv{tabular}
\makesavenoteenv{table}

\usepackage[margin=1in]{geometry}

\definecolor{crimsonglory}{rgb}{0,0,0}
\DeclarePairedDelimiter\ceil{\lceil}{\rceil}
\DeclarePairedDelimiter\floor{\lfloor}{\rfloor}

 \newtheorem{theorem}{Theorem}[section]
 
 \newtheorem{lemma}[theorem]{Lemma}
 \newtheorem{proposition}[theorem]{Proposition}

\makeatletter
\def\GrabProofArgument[#1]{ #1: \egroup\ignorespaces}
\def\proof{\noindent\textbf\bgroup Proof%
	\@ifnextchar[{\GrabProofArgument}{. \egroup\ignorespaces}}

\makeatother

\usetikzlibrary{arrows,shapes,snakes,automata,backgrounds,petri,calc}
\usepackage[latin1]{inputenc}

\newcommand{\E}{\mathbb E}
\newcommand{\Prb}{\mathbb P}
\newcommand{\OPT}{\mathsf{OPT}}
\newcommand{\LP}{\mathsf{LP}}
\newcommand{\dyad}{\mathsf{dyad}}
\newcommand{\Bin}{\mathsf{Bin}}
\newcommand{\MC}{\mathsf{MC}}
\newcommand{\CoverRow}{\textsf{CoverRow}}
\newcommand{\MultiCoverRow}{\textsf{MultiCoverRow}}
\newcommand{\ind}{\mathbf 1}
\newcommand{\polylog}{\operatorname{polylog}}
\renewcommand{\ceil}[1]{\left\lceil#1\right\rceil}
\renewcommand{\floor}[1]{\left\lfloor#1\right\rfloor}

\newcommand*{\leanrepository}{https://github.com/hajiaghayi/Online-Survivable-Network}

\newcounter{claudecnt}

\newif\ifshowclaudeedits
\showclaudeeditstrue 

\newcommand{\konote}[1]{}

\usepackage[normalem]{ulem} 

\setlist{itemsep=0.2em,topsep=0.35em}
\allowdisplaybreaks[2]
\hypersetup{pdftitle={Online Survivable Network Design with Arbitrarily Small Connectivity Loss},pdfauthor={Mohammad Hajiaghayi and Saeed Seddighin}}

\title{Online Survivable Network Design with Arbitrarily Small Connectivity Loss}

\author{
 Mohammad Hajiaghayi\thanks{\texttt{hajiaghayi@gmail.com}}
 \and
 Saeed Seddighin\thanks{\texttt{saeedreza.seddighin@gmail.com}}
}

\SetCommentSty{mycommfont}

\SetKwInput{KwInput}{Input}                
\SetKwInput{KwOutput}{Output}
\begin{document}
	\newcommand{\ignore}[1]{}
\renewcommand{\theenumi}{(\roman{enumi}).}
\renewcommand{\labelenumi}{\theenumi}
\sloppy
\date{}

\maketitle

\thispagestyle{empty}

\begin{abstract}
In the online survivable network design problem, we are given a graph
with edge costs in advance, and connectivity requests arrive one at a
time. Each request consists of a pair of vertices $(s,t)$ and a
requirement $k$, and upon its arrival the algorithm must irrevocably buy
edges so that $s$ and $t$ are connected by $k$ edge-disjoint paths. The
goal is to keep the total cost of the purchased edges close to the cost
of an optimal offline solution.

Gupta, Krishnaswamy, and Ravi~\cite{GKR12} give an
$O(k\log^3 n)$-competitive algorithm for this problem, where $k$ is the
maximum requirement, without relaxing the connectivity requirement.
Dehghani, Ehsani, Hajiaghayi, Liaghat, and Seddighin~\cite{DEHLS18}
obtain a polylogarithmic competitive ratio when the connectivity
requirement may be relaxed by a factor of $2+\varepsilon$. In this work,
we show that relaxing the connectivity requirement by a factor of only
$1+\varepsilon$ already suffices for a polylogarithmic competitive
ratio.

\end{abstract}
\section{Introduction}
\label{sec:introduction}

Survivable network design is one of the central problems in the study of
fault-tolerant networks. We are given a graph whose edges have costs, and
a collection of vertex pairs, each with a connectivity requirement. The
goal is to buy a cheapest set of edges such that every pair is connected
by as many edge-disjoint paths as it requires. Such a network keeps every
requested pair connected even after a number of edge failures, and this
is exactly the property that the requirement encodes. In the offline
setting, Jain~\cite{Jain01} gives a celebrated factor-two approximation
for this problem.

In this work, we study the online variant of survivable network design.
Here the network is known in advance, but the pairs that need protection
arrive over time. Upon the arrival of a request, consisting of a pair
$(s,t)$ and a requirement $k$, the algorithm must buy enough edges to
provide $k$ edge-disjoint $s$--$t$ paths. Every bought edge remains
available for all future requests, and its cost is paid only once. The
decisions are irrevocable: the algorithm can never sell an edge. Its
performance is measured against the cheapest network that serves all
requests seen so far. When all requirements are equal to one, the problem
reduces to the classic online Steiner tree and Steiner forest problems,
studied by Imase and Waxman~\cite{IW91} and Berman and
Coulston~\cite{BC97}, among others. Larger requirements make the problem
considerably harder: the algorithm must now preserve several independent
routes for every pair, while it makes irrevocable decisions about which of
these routes are shared between different pairs.

Gupta, Krishnaswamy, and Ravi~\cite{GKR12} give a randomized
$O(k\log^3 n)$-competitive algorithm for online survivable network design
that provides the full requirement $k$ to every pair. (When requirements
differ between requests, $k$ denotes the largest one.) Their method uses
spanning trees and an online covering problem whose explicit
representation has size $m^{O(k)}$. In the same work, they show that a
logarithmic lower bound holds already for rooted requests of requirement
two~\cite[Appendix A]{GKR12}. The competitive ratio of~\cite{GKR12} grows
linearly with the requirement, and a natural way to avoid this growth is
to relax the amount of connectivity that the algorithm provides.
Dehghani, Ehsani, Hajiaghayi, Liaghat, and Seddighin~\cite{DEHLS18}
follow this direction. When all requirements are equal to $k$, their
algorithms provide only $k/2$ edge-disjoint paths for every pair and, in
exchange, achieve competitive ratios that do not depend on $k$: the
competitive ratio is $O(\log n)$ when all requests share a common
endpoint, and $O(\log^2 n)$ for arbitrary pairs.

These results raise a natural question: is it necessary to lose half of
the connectivity in order to obtain a competitive ratio independent of
$k$? In this work, we show that it is not, and in fact every prescribed
relative slack is sufficient. Fix any $0<\varepsilon\le1$. For a request
with requirement $k$, our algorithm always provides
$\ceil{k/(1+\varepsilon)}$ edge-disjoint paths. We compare its expected
cost with the cheapest network that provides \emph{all} $k$ paths, and we
show that its competitive ratio is polylogarithmic times $1/\varepsilon$.
More precisely, our results are as follows.
\begin{itemize}
 \item When all requirements are equal to $k$, the competitive ratio is
 $O(\log^3 n/\varepsilon)$. If, in addition, $k$ is at least a constant
 fraction of $n$, the competitive ratio drops to $O(\log n/\varepsilon)$.
 \item When requirements differ, every request still receives a
 $1/(1+\varepsilon)$ fraction of its own requirement, rounded up, and the
 competitive ratio is $O(\log n\log^2 m/\varepsilon)$. This competitive
 ratio does not grow with the largest requirement, and the algorithm does
 not need to know this requirement in advance.
\end{itemize}

Our algorithms run in time polynomial in the input size, regardless of
the value of $k$. They are randomized, but they never fail to provide the
promised connectivity: only their cost is random. The cost guarantee
holds in expectation at every point of a request sequence that is fixed
in advance. In addition, when all requirements are equal to $k$, we show
how the cost changes if the algorithm provides an arbitrary number
$q\le k$ of paths. It turns out that the quantity that governs the cost
is the \emph{slack} $k-q+1$. Whether exact connectivity admits a
competitive ratio polylogarithmic in $n$ remains an open question.

\Cref{tab:comparison} summarizes our results along with the previous
work. We formally state our results in \cref{sec:results} and give an
overview of our techniques in the rest of \cref{sec:approach}.

\begin{table}[t]
\centering
\small
\renewcommand{\arraystretch}{1.25}
\begin{tabular}{@{}>{\raggedright\arraybackslash}p{0.23\linewidth}>{\raggedright\arraybackslash}p{0.21\linewidth}>{\raggedright\arraybackslash}p{0.14\linewidth}>{\raggedright\arraybackslash}p{0.33\linewidth}@{}}
\toprule
Result & Requests & Loss & Competitive ratio \\
\midrule
Gupta et al.~\cite{GKR12} & General, maximum $k$ & None & $O(k\log^3 n)$ \\
Dehghani et al.~\cite{DEHLS18} & Uniform, rooted & $2$ & $O(\log n)$ \\
Dehghani et al.~\cite{DEHLS18} & Uniform, arbitrary pairs & $2$ & $O(\log^2 n)$ \\
This paper & Uniform, arbitrary pairs & $1+\varepsilon$ & \eqref{eq:headline}; $O(\varepsilon^{-1}\log n)$ if $k=\Omega(n)$ \\
This paper & Nonuniform, arbitrary pairs & $1+\varepsilon$ & $O(\varepsilon^{-1}\log(n+1)\log^2(m+1))$ \\
\bottomrule
\end{tabular}
\caption{The results of this paper along with previous work. All cost
bounds are with respect to the optimum for the full requirements, and the
guarantees of this paper hold even against the capacitated cut LP. The
loss is the factor by which the connectivity requirement is relaxed. Our
algorithms are randomized against oblivious arrivals, whereas the cited
algorithms with loss two are deterministic.}
\label{tab:comparison}
\end{table}

\subsection{Related work}
\label{sec:related}

Our improvement over previous work is in the permitted connectivity loss,
in the treatment of arbitrary requirements, and in the dependence on
connectivity within the new tradeoff. We do not, however, improve every
earlier special case. For example, the deterministic $O(\log n)$
competitive ratio of~\cite{DEHLS18} for rooted uniform requests remains
better than ours at loss two in some parameter ranges.

For nonuniform requirements, the bicriteria framework of~\cite{DEHLS18}
incurs a factor $\log^{1+\delta}k$ in the maximum requirement $k$ and a
loss of $2+\delta$, for a constant $\delta>0$. In contrast, our
nonuniform algorithm handles all requirements in a single covering
process. Instead of partitioning the requests into requirement classes,
its analysis works with residual multiplicities. Naor, Umboh, and
Williamson~\cite{NUW19} improve the special case of maximum requirement
two through online weighted tree augmentation.

Online covering and primal--dual growth are standard tools in online
algorithms~\cite{AAABN09,BN09}, and our algorithms build on them. Online
covering has also been combined with packing constraints in network
design: Dehghani, Ehsani, Hajiaghayi, Liaghat, R\"acke, and
Seddighin~\cite{DEHLRS16} obtain online algorithms for edge-weighted
degree-bounded Steiner forest through a new online algorithm for mixed
packing/covering integer programs. Beyond adversarial arrivals, Dehghani et al.~\cite{DEHLS18}
also analyze the greedy algorithm for online survivable network design
when the requests are drawn from a distribution, and stochastic online
inputs have been studied for the $k$-server problem as
well~\cite{DEHLS17}. Borst,
Eli\'a\v{s}, and Venzin~\cite{BEV25} use a model of semi-adaptive
adversaries as a key ingredient of their algorithm for online
node-weighted Steiner forest. Our analysis faces a related issue: the
minimum cut that our algorithm repairs at any moment depends on its
previous random purchases. We deal with this issue by proving one
exponential-moment inequality for each member of a fixed family of
shores. The new quantitative ingredients of our analysis are the
multiplicity discount, the coordinate cap, and the connectivity-rank
bound on the number of repairs; we explain each of them in
\cref{sec:approach}.

Connectivity augmentation on a fixed base graph has also seen recent
progress. Garg and Subramanian~\cite{GS26} study the online augmentation
of a given $k$-edge-connected graph, where each arriving pair must become
$(k+1)$-edge-connected. They obtain deterministic competitive ratios of
$O(\log n)$ and, for weighted links, $O(\log^2 n)$. Garg, Hommelsheim,
and Lindermayr~\cite{GHL26} obtain a deterministic $O(\log n)$ bound for
weighted links when pairs arrive, and Hathcock, Subramanian, Umboh, and
Zlatin~\cite{HSUZ26} obtain nearly logarithmic weighted bounds in the
model where cuts arrive. All these results start from a fixed graph that
is globally $k$-edge-connected. In our setting, by contrast, the bought
graph grows with the requests, and the Steiner vertices that are not
requested need not have high connectivity.

Finally, our algorithms use random spanning trees of the input graph with
small expected stretch. We emphasize that we need trees that consist of
input edges: a distribution over arbitrary tree metrics, such as the
embedding of Fakcharoenphol, Rao, and Talwar~\cite{FRT04}, does not by
itself provide the physical cycles that our algorithm buys. We discuss
the spanning-tree samplers that we use at the end of \cref{sec:results}.

\subsection{Preliminaries}
\label{sec:model}

\paragraph{Graphs and cuts.}
The input is a known loopless undirected multigraph $G=(V,E)$ with
$n=|V|\ge2$ vertices and $m=|E|$ edges. Parallel edges are allowed, and
we call the set of edges with the same two endpoints a \emph{parallel
bundle}. Each edge $e$ has a nonnegative rational cost $c_e$, and for a
set of edges $F$ we write $c(F)=\sum_{e\in F}c_e$. Each edge has unit
capacity and can be bought at most once. We identify a set of edges with
the spanning subgraph of $G$ that it forms. In particular, connectivity
in a set of bought edges $H$ refers to the graph $(V,H)$. For a vector
$y\in\mathbb R^E$ and a set $F\subseteq E$, we write
$y(F)=\sum_{e\in F}y_e$ and $c^\top y=\sum_{e\in E}c_ey_e$.

A \emph{shore} is a set of vertices $S\subseteq V$. For a subgraph $J$ of
$G$, we denote by $\delta_J(S)$ the set of edges of $J$ with exactly one
endpoint in $S$. A shore $S$ \emph{separates} two vertices $u$ and $v$ if
it contains exactly one of them. A $u$--$v$ \emph{separator} in $J$ is a
set of edges of $J$ whose removal leaves no $u$--$v$ path. Notice that a
separator is a set of edges whereas a shore is a set of vertices; in our
parity construction, these two objects need not carry the same
information. The \emph{connectivity} $\lambda_J(u,v)$ is the minimum size
of a $u$--$v$ separator in $J$, and a separator of this size is a
\emph{minimum} separator. By the edge version of Menger's theorem,
$\lambda_J(u,v)$ is also the maximum number of edge-disjoint $u$--$v$
paths in $J$. A \emph{minimum-cut shore} for $u$ and $v$ in $J$ is a
shore $S$ that separates $u$ and $v$ and satisfies
$|\delta_J(S)|=\lambda_J(u,v)$. All connectivity computations use unit
capacities on the edges; costs are never used as capacities.

\paragraph{Requests.}
Requests $(s_i,t_i,r_i)$ arrive one at a time, for $i=1,2,\ldots$. Each
request has distinct endpoints $s_i\ne t_i$ and a positive integer
requirement $r_i$. We assume that the requests are feasible in $G$, that
is, $\lambda_G(s_i,t_i)\ge r_i$; thus every requirement is at most $m$.
Disconnected inputs can be treated componentwise, and therefore we
describe the algorithms for one connected component. The requests are
\emph{rooted} if all of them share a common endpoint. The first $i$
requests form the \emph{prefix}~$i$. When we describe an algorithm, we
write $(u,v)$ for the pair of the request being processed. Requirements
are \emph{uniform} if all of them are equal to a common value $k$, known
in advance; in this case, we write $k$ instead of $r_i$. In general, $k$
refers to the largest requirement, which our nonuniform algorithm does
not need to know. An online algorithm maintains a set $H$ of bought
edges, which only grows. We denote this set by $H_i$ after request $i$ is
processed; the total cost paid by the algorithm up to that point is
$c(H_i)$.

\paragraph{Connectivity loss and slack.}
For a parameter $0<\varepsilon\le1$ (a positive rational when supplied as
input), the algorithm must ensure that
$\lambda_{H_i}(s_j,t_j)\ge\ceil{r_j/(1+\varepsilon)}$ for every $j\le i$.
We say that the algorithm violates the connectivity requirement by a
factor of $1+\varepsilon$, and we call $1+\varepsilon$ its
\emph{connectivity loss}. For example, providing only half of each
requirement corresponds to loss $2$. For uniform requirements, we also
study an arbitrary integer target $1\le q\le k$, where every requested
pair must have connectivity at least $q$ in $H$. We call
\[
 \sigma=k-q+1
\]
the \emph{slack}. Our main case is $q=\ceil{k/(1+\varepsilon)}$.

\paragraph{Benchmarks.}
A request sequence is \emph{oblivious} if it is fixed independently of
all random choices of the algorithm. The expectations $\E[\cdot]$ and the
probabilities $\Prb[\cdot]$ are over the random choices of the
algorithm. Notice that the cuts that the algorithm computes internally
may still depend on its entire purchase history.

For a prefix $i$, let $\mathcal C_i$ be the family of shores that
separate at least one requested pair $(s_j,t_j)$ with $j\le i$. When all
requirements are equal to $k$, we define
\begin{equation}
 \LP_k(i)=\min\left\{\sum_{e\in E}c_e y_e:\
 0\le y_e\le1,\quad
 y(\delta_G(S))\ge k\ \text{for every }S\in\mathcal C_i\right\}.
 \label{eq:uniform-lp}
\end{equation}
Its integral counterpart, in which $y_e\in\{0,1\}$, is $\OPT_k(i)$; this
is the cost of the cheapest set of edges that gives every requested pair
connectivity $k$. The bounds $y_e\le1$ reflect the unit capacities, and
they are essential to our analysis. For nonuniform requests, $\LP(i)$ and
$\OPT(i)$ are defined similarly, except that a shore that separates
$(s_j,t_j)$ must be crossed by a mass of at least $r_j$. In every case,
$\LP(i)\le\OPT(i)$. An algorithm is \emph{$\alpha$-competitive} if
$\E[c(H_i)]\le\alpha\,\OPT(i)$ for every prefix $i$ of every oblivious
sequence. All our bounds are stated against $\LP$, which is a stronger
guarantee.

Throughout our cost analysis, we fix a finite prefix of the request
sequence and one feasible vector $y$ of its cut LP for the full
requirements. We refer to $y$ as the \emph{comparator}. The comparator is
independent of the sampled tree and of all purchase coins, and it is used
only in the analysis: the algorithm never needs to know the future
requests, an optimum solution, or the value of $\LP$.

\paragraph{Spanning trees and stretch.}
Let $T\subseteq G$ be a spanning tree. For two vertices $u$ and $v$, we
denote by $P_T(u,v)$ the unique $u$--$v$ path in $T$, and by
$d_T(u,v)=c(P_T(u,v))$ its cost; tree edges keep their original costs.
For an edge $e$ with endpoints $u$ and $v$, we write $P_T(e)=P_T(u,v)$
and $d_T(e)=d_T(u,v)$. An edge $e\notin T$ is a \emph{chord}, and its
\emph{fundamental cycle} is $C_e=\{e\}\cup P_T(e)$. A polynomial-time
spanning-tree sampler has \emph{stretch} $\rho\ge1$ if it outputs a
random spanning tree $T\subseteq G$ such that
\begin{equation}
 \E_T[d_T(u,v)]\le \rho\, c_e\qquad\text{for every edge } e=uv\in E.
 \label{eq:stretch}
\end{equation}
The stretch $\rho$ may depend on $n$. We say that a spanning tree $T$ is
\emph{sampled} if it is drawn from such a sampler; all later random
choices of the algorithm are independent of $T$. In our main bounds, we
use the sampler of Hajiaghayi~\cite{Hajiaghayi26} with
$\rho=O(\log(n+1))$, while our more detailed tradeoff keeps $\rho$
explicit. Our direct-cut algorithm does not use a spanning tree at all.

\paragraph{Other notation.}
Throughout the paper, $\log$ is to an arbitrary fixed base, $\ln$ is the
natural logarithm, $\exp(x)=e^x$, and $\ind[\cdot]$ is the indicator of
an event. The notation $\polylog n$ means $(\log n)^{O(1)}$, and
$O_\alpha(\cdot)$ hides a factor that depends only on the parameter
$\alpha$.

\section{Our Results and Techniques}
\label{sec:approach}

In this section, we formally state our results and explain the main
ideas behind them. We state the results in \cref{sec:results}. In
\cref{sec:overview-discount,sec:overview-cap,sec:overview-adaptive}, we
describe a simple algorithm, which we call the \emph{direct-cut
algorithm}, and use it to introduce the three ingredients that all our
algorithms share. The direct-cut algorithm already achieves a
logarithmic competitive ratio when $k$ is linear in $n$. In
\cref{sec:overview-parity}, we explain how a spanning tree allows us to
obtain a polylogarithmic competitive ratio for every $k$; this is our
\emph{parity-cover algorithm}. Finally, in
\cref{sec:overview-nonuniform}, we explain how the same approach handles
nonuniform requirements.

\subsection{Our results}
\label{sec:results}

Our first result concerns uniform requirements. It shows that, for every
$\varepsilon>0$, a connectivity loss of $1+\varepsilon$ is enough to
obtain a polylogarithmic competitive ratio.

\begin{theorem}[Uniform requirements]
\label{thm:headline}
For every $0<\varepsilon\le1$ and every known common requirement $k$,
there is a randomized polynomial-time online algorithm that always
provides edge connectivity $q=\ceil{k/(1+\varepsilon)}$ for all arrived
pairs. For every fixed prefix $i$ of an oblivious request sequence, it
satisfies
\begin{equation}
 \E[c(H_i)]\le
 O\!\left(\frac{\log(n+1)}{\varepsilon}
 \min\left\{\log(n+1)\log\left(2+\frac nk\right),\,
                   1+\frac nk\right\}\right)\LP_k(i).
 \label{eq:headline}
\end{equation}
This holds for arbitrary multigraphs. In particular, if $k\ge\alpha n$
for a fixed $\alpha>0$, the competitive ratio is
$O_\alpha(\varepsilon^{-1}\log(n+1))$, and this bound does not depend on
the spanning-tree sampler.
\end{theorem}

The algorithm of \cref{thm:headline} is one of our two algorithms, the
parity-cover algorithm or the direct-cut algorithm. The graph, $k$, and
$\varepsilon$ determine which of the two is used, and therefore the
choice is made before any request arrives. One factor $\log(n+1)$ in
\eqref{eq:headline} comes from discarding all but the $k$ cheapest edges
of each parallel bundle, a reduction that preserves both benchmarks.
Before this reduction, we prove the following more precise tradeoff
between the provided connectivity and the cost. It holds for every
target $q\le k$, not only for $q=\ceil{k/(1+\varepsilon)}$.

\begin{theorem}[The full connectivity--cost tradeoff]
\label{thm:tradeoff}
Fix integers $1\le q\le k$ and a common full requirement $k$. The
parity-cover algorithm, run with a spanning-tree sampler of stretch
$\rho$, provides connectivity $q$ at expected cost at most
\begin{equation}
 O\!\left(\rho\left[\frac1k+
 \frac{q-1}{k-q+1}
 \log\left(2+\frac nq\right)
 \log\left(1+\frac mq\right)\right]\right)\LP_k(i).
 \label{eq:parity-tradeoff}
\end{equation}
For $q=1$, its bound is exactly $\rho\LP_k(i)/k$. The direct-cut
algorithm provides the same target at expected cost at most
\begin{equation}
 O\!\left(\frac{n+q}{k-q+1}
 \log\left(1+\frac{m}{n+q}\right)\right)\LP_k(i).
 \label{eq:direct-tradeoff}
\end{equation}
Both algorithms run in polynomial time, are feasible on every execution,
and satisfy the displayed bounds at every fixed prefix of an oblivious
sequence.
\end{theorem}

The denominator $k-q+1$ in both bounds is the slack $\sigma$. In other
words, the cost of both algorithms improves as the target $q$ moves away
from the full requirement $k$. Notice that at $q=k$, the parity bound
still has a factor $k-1$. Thus, at this endpoint, \cref{thm:tradeoff}
gives a polynomial-time algorithm for exact connectivity whose
competitive ratio grows linearly with $k$; it does not give an exact
$o(k)\polylog n$ bound in general.

Our second main result concerns arbitrary requirements. Here, too, a
connectivity loss of $1+\varepsilon$ is enough for a polylogarithmic
competitive ratio, and the algorithm does not need to know the
requirements of future requests.

\begin{theorem}[Nonuniform requirements]
\label{thm:nonuniform}
For arbitrary online requests $(s_i,t_i,r_i)$ and a fixed
$0<\varepsilon\le1$, there is a randomized polynomial-time algorithm that
always provides edge connectivity $\ceil{r_i/(1+\varepsilon)}$ to each
request and satisfies
\begin{equation}
 \E[c(H_i)]\le
 O\!\left(\frac{\log(n+1)\log^2(m+1)}{\varepsilon}\right)\LP(i)
 \label{eq:nonuniform-main}
\end{equation}
at every fixed prefix of an oblivious sequence. No upper bound on future
requirements is needed. In particular, the competitive ratio has no
additional factor that depends on the maximum requirement.
\end{theorem}

Our theorems are stated for requests between arbitrary pairs. They also
include the original formulation of the problem, in which terminals
arrive one by one and every two terminals must be connected. To see this,
take the first terminal as a root and request a pair between the root and
every later terminal. Providing connectivity $q$ from every terminal to
the root provides connectivity $q$ between every two terminals, since a
cut that separates two terminals separates at least one of them from the
root.

\paragraph{Spanning-tree samplers.}
The logarithmic spanning-tree theorem of Hajiaghayi~\cite{Hajiaghayi26}
gives a polynomial-time sampler with stretch $\rho=O(\log(n+1))$, and
this is the sampler that we use in \cref{thm:headline,thm:nonuniform}.
Consequently, both uniform requests on arbitrary multigraphs and
nonuniform requests on simple graphs admit a competitive ratio of
$O(\varepsilon^{-1}\log^3(n+1))$. For uniform requests, the stronger
bound \eqref{eq:headline}, which depends on $k$, applies.

This stretch matches the $\Omega(\log n)$ lower bound of Alon, Karp,
Peleg, and West~\cite{AKPW95}, which holds already for the average
stretch of a single spanning tree. We use the result
of~\cite{Hajiaghayi26} as an external input. The proof of
\cref{thm:tradeoff} works with any sampler that satisfies
\eqref{eq:stretch}. For completeness, \cref{app:trees} also shows how the
published average-stretch bound of Abraham and Neiman~\cite{AN19} yields
such a sampler. If we use the Abraham--Neiman bound instead
of~\cite{Hajiaghayi26}, every occurrence of $\rho$ in our bounds costs an
additional factor $O(\log\log n)$.

\subsection{Deficient cuts and the multiplicity discount}
\label{sec:overview-discount}

Fix a common full requirement $k$, and suppose that we only need to
provide connectivity $q\le k$. Recall that $\sigma=k-q+1$ denotes the
slack. Our algorithms follow the general framework of online covering:
whenever the current request is not served, the algorithm finds a
violated constraint, which we call a \emph{row}, and buys edges until the
row is hit. The quality of such an algorithm depends on how much the
benchmark must invest in each row, and our first observation is that,
thanks to the slack, the benchmark covers every row many times over.

Let us see this in the simplest algorithm, the direct-cut algorithm.
Suppose that the current request $(u,v)$ has connectivity $\ell<q$ in the
bought graph $H$, and let $S$ be a minimum-cut shore for $u$ and $v$ in
$H$. Exactly $\ell$ bought edges cross $S$, and the algorithm exposes the
row
\[
 A=\delta_G(S)\setminus H
\]
of unbought edges that cross $S$. Buying any edge of $A$ increases the
size of this particular cut. Now consider any network $F$ that provides
the full connectivity $k$ to all requested pairs. Since $S$ separates $u$
and $v$, at least $k$ edges of $F$ cross $S$, and at most $\ell$ of them
are already bought. Therefore,
\begin{equation}
 |F\cap A|\ge k-\ell\ge \sigma.
 \label{eq:overview-multiplicity}
\end{equation}
In other words, the benchmark contains at least $\sigma$ witnesses for
every row, while the algorithm needs to buy only one of them. The same
calculation applies to the comparator $y$: since $y_e\le1$ for every
edge, we have $y(A)\ge k-|\delta_H(S)|\ge k-\ell\ge\sigma$. Thus the
scaled vector $y/\sigma$ is a fractional cover of \emph{every} row that
the algorithm may expose. This discount by a factor of $\sigma$ is the
source of our improvement as the target falls below $k$. Notice that the
argument uses all available witnesses, even though the algorithm only
needs one of them to make progress.

\subsection{Capped fractional growth}
\label{sec:overview-cap}

The rows are not given in advance. The algorithm finds one deficient cut
at a time, forms its row, and processes it as follows. A fractional
variable $x_e$ records the investment in object $e$; in the direct-cut
algorithm the objects are edges, and their prices are $d_e=c_e$. As long
as the row $A$ is not hit and $x(A)<1$, the algorithm performs a
cost-scaled multiplicative increase with a small additive seed:
\[
 \Delta x_e=\frac{\gamma}{d_e}\left(x_e+\frac1m\right)
 \qquad\text{for every }e\in A,
\]
where $\gamma$ is a small step size. Each increase also gives the object
a fresh chance to be bought, with probability proportional to
$a\,\Delta x_e$. Here $a$ is an \emph{amplification} parameter, which, as
we explain shortly, is determined by the number of different shores that
may witness a repair. Once an object is bought, its coordinate is frozen.

The crucial additional rule is deterministic: the algorithm buys an
object as soon as $x_e$ reaches $1/a$. Thus, the fractional process never
invests substantially beyond the scale at which the rounding should have
selected the object anyway. Allowing a constant overshoot, we have
$x_e\le3/(2a)$ at all times. The usual multiplicative-growth calculation
then bounds the total dual load on an object by
\[
 O\bigl(d_e\log(1+m/a)\bigr)
\]
instead of $O(d_e\log m)$. Combining this with
\eqref{eq:overview-multiplicity} bounds the fractional cost by
$O(\log(1+m/a))\, d^\top y/\sigma$. Random purchases and threshold
purchases multiply this bound by $O(a)$, and therefore the total expected
cost of the purchases is
\[
 O\!\left(\frac{a\log(1+m/a)}\sigma\right)d^\top y.
\]
It remains to show that a small amplification $a$ suffices. The
difficulty is that every row depends on the previous random purchases of
the algorithm.

\subsection{Adaptive rows and one exponential moment per shore}
\label{sec:overview-adaptive}

If a row reaches mass one before any of its objects is bought, the
algorithm forces a purchase of a cheapest object of the row. We call this
a \emph{forced repair}. The expected cost of forced repairs is small as
long as they are unlikely, and we need to show this for rows that the
algorithm chooses adaptively, based on its previous coins.

The key idea is to analyze shores rather than rows. Fix a shore $S$, and
let $\mathcal U_S$ be the set of objects that cross $S$. We track two
quantities: the fractional mass $x(\mathcal U_S)$ and the number
$|I\cap\mathcal U_S|$ of selected objects that cross $S$, where $I$ is the
set of all selected objects. Fractional investment increases the first
quantity, while actual purchases increase the second. With
$\beta=1-e^{-1}$, we show that the process
\begin{equation}
 \psi_S=\exp\bigl(\beta a\,x(\mathcal U_S)-|I\cap\mathcal U_S|\bigr)
 \label{eq:overview-potential}
\end{equation}
is a nonnegative supermartingale. Intuitively, since the coins are fresh
and independent, the expected decrease caused by purchases cancels the
increase caused by fractional growth, and deterministic purchases can
only decrease $\psi_S$. This calculation is conditional on the entire
history, and therefore it allows the next row to be chosen adaptively.

Now consider a forced repair of a row $A$ whose shore is $S$. At this
moment, $x(\mathcal U_S)\ge x(A)\ge1$. Moreover, at most $q-1$ selected
objects cross $S$; in the direct-cut algorithm, for example, the selected
objects that cross $S$ are bought edges of a cut of size $\ell\le q-1$.
Hence $\psi_S\ge\exp(\beta a-q+1)$ at every forced repair. By a maximal
inequality for nonnegative supermartingales, this happens with
probability at most $\exp(-\beta a+q-1)$ for a fixed shore. Let $\nu$ be
an upper bound on the number of shores, and let $\tau$ be an upper bound
on the number of rows that the algorithm processes. A union bound over the \emph{fixed} family
of shores controls all adaptive repairs at once: if
\[
 \nu\exp(-\beta a+q-1)\le\frac1{2\tau},
\]
then the expected cost of forced repairs is at most
$d^\top y/(2\sigma)$. We never enumerate the shores; only the logarithm of
their number enters the amplification, and hence the cost.

For the direct-cut algorithm, there are fewer than $\nu=2^{n-1}$ shores.
To bound the number of rows, we use the following potential. For every
level $h=1,\ldots,q$, consider the partition of the vertices into
classes of $h$-edge-connectivity in the bought graph. The first edge
bought while processing a row joins two vertices whose connectivity is
below $q$, and therefore merges two classes at some level. Summed over
all levels, there are at most $q(n-1)$ such mergers. This gives
$\tau\le q(n-1)$, a deterministic bound that depends on the number of
vertices and on the connectivity, but not on the number of parallel
edges. Consequently, an amplification of $a=\Theta(n+q)$ suffices, and
the cost bound becomes
\[
 O\!\left(\frac{n+q}{k-q+1}
       \log\left(1+\frac m{n+q}\right)\right)\LP_k.
\]
The factor $n$ comes from the logarithm of the number of shores, and the
denominator comes from the number of witnesses. For
$q=\ceil{k/(1+\varepsilon)}$ and $k=\Omega(n)$, these two factors cancel
up to $O(1/\varepsilon)$. This explains the logarithmic competitive ratio
of \cref{thm:headline} when $k$ is linear in $n$, without any
tree-embedding factor. For small $k$, however, the ratio $n/k$ is large,
and we need a much smaller family of shores.

\subsection{Parity shores: compressing the shores with a spanning tree}
\label{sec:overview-parity}

To reduce the number of shores, the parity-cover algorithm samples one
spanning tree $T$ at the beginning. On a request $(u,v)$, it first buys
the tree path $P_T(u,v)$. All remaining decisions are purchases of
fundamental cycles
\[
 C_e=\{e\}\cup P_T(e),\qquad e\notin T,
\]
at prices $d_e=c_e+d_T(e)$. These prices may overcount edges that are
already bought, which only hurts the algorithm. On the other hand, they
give a fixed price to each of at most $m$ objects, and the expected
stretch of the tree later relates these prices to the original costs of
the comparator.

Suppose that the bought graph $H$ still has a minimum $u$--$v$ separator
$Q$ of size $\ell<q$. Root $T$ at a reference vertex, and label each
vertex by the parity of the number of edges of $Q\cap T$ on its tree path
from the root. Let $S_Q$ be the set of vertices with odd labels; see
\cref{fig:parity}. The bought tree path from $u$ to $v$ crosses $Q$ an
odd number of times, so $S_Q$ separates the requested pair. The
algorithm exposes the row
\[
 A_Q=\delta_G(S_Q)\setminus Q,
\]
which consists of chords whose fundamental cycles meet $Q$ an odd number
of times. Buying such a cycle destroys $Q$ as a separator. The reason is
elementary: if $Q$ remained a separator, then its minimality would force
it to be a cut of the enlarged graph, whereas every cycle crosses every
cut an even number of times.

\begin{figure}[t]
\centering
\begin{tikzpicture}[x=1cm,y=1cm,>=Latex,font=\small]
\foreach \x/\v/\lab in {0/v0/0,2/v1/1,4/v2/1,6/v3/0,8/v4/0,10/v5/1}{
 \node[circle,draw=black,fill=white,minimum size=6mm,inner sep=0pt] (\v) at (\x,0) {$\lab$};
}
\draw[very thick,BrickRed,dashed] (v0)--(v1);
\draw[thick] (v1)--(v2);
\draw[very thick,BrickRed,dashed] (v2)--(v3);
\draw[thick] (v3)--(v4);
\draw[very thick,BrickRed,dashed] (v4)--(v5);
\node[BrickRed] at (1,0.55) {$Q\cap T$};
\node[BrickRed] at (5,0.55) {$Q\cap T$};
\node[BrickRed] at (9,0.55) {$Q\cap T$};
\draw[very thick,MidnightBlue] (v0.south) to[out=-65,in=-115] node[below=6pt,pos=.5] {$e\in A_Q$: buy $C_e=P_T(e)\cup\{e\}$} (v5.south);
\node[above=11mm] at (v0) {$u$};
\node[above=11mm] at (v5) {$v$};
\node[align=center] at (5,1.35) {Parity labels flip exactly on $Q\cap T$.};
\end{tikzpicture}
\caption{The parity labels flip exactly on the edges of $Q\cap T$. A
chord whose endpoints have different labels has a fundamental cycle that
meets $Q$ an odd number of times. The tree path shown has three edges of
$Q$; only their parity matters. Notice that the shore $S_Q$ need not be a
minimum-cut shore of the bought graph.}
\label{fig:parity}
\end{figure}

The multiplicity discount survives this construction. Every full-$k$
solution has at least $k$ edges across $S_Q$, and removing $Q$ excludes
at most $\ell$ of them; hence the comparator satisfies
$y(A_Q)\ge k-\ell\ge\sigma$, exactly as in
\eqref{eq:overview-multiplicity}.

The gain is in the number of shores. The parity shore $S_Q$ depends only
on $Q\cap T$, which is a set of at most $q-1$ of the $n-1$ tree edges.
Therefore, there are only
\[
 \nu_T=\sum_{j=1}^{\min\{q-1,n-1\}}\binom{n-1}{j}
\]
possible nontrivial parity shores, and
$\ln\nu_T=O(q\log(2+n/q))$. Notice that many different rows may have the
same parity shore, since they may exclude different bought chords. This
is why we track the potential \eqref{eq:overview-potential} for shores
rather than for rows. At a forced repair, every selected chord that
crosses $S_Q$ belongs to $Q$, since otherwise its cycle would already
have destroyed the separator. Thus at most $q-1$ selected chords cross
$S_Q$, and the argument of \cref{sec:overview-adaptive} applies
unchanged.

The number of rows is bounded by a similar potential, now in the virtual
graph $T\cup I$ and over the levels $h=2,\ldots,q$; this gives
$\tau_T\le(q-1)(n-1)$. Consequently, we can take
$a=O(q\log(2+n/q))$. Together with $a\ge q$, the cap gives the product
\[
 \log\left(2+\frac nq\right)
 \log\left(1+\frac mq\right)
\]
in \eqref{eq:parity-tradeoff}. Finally, the requested tree paths cost at
most a $1/k$ fraction of the total tree distance of the comparator, since
every used tree edge defines a cut that the comparator crosses at least
$k$ times. Averaging the cycle charges and the path charges over $T$
contributes the factor $\rho$.

\Cref{fig:two-algorithms} summarizes how the two algorithms compare.
Since both of their bounds are known before the first request arrives,
the algorithm of \cref{thm:headline} simply runs the one with the smaller
bound.

\begin{figure}[t]
\centering
\begin{tikzpicture}[font=\small,box/.style={draw=MidnightBlue,rounded corners=2pt,fill=MidnightBlue!3,align=center,text width=5.7cm,minimum height=2.4cm,inner sep=8pt}]
\node[box] (parity) at (0,0) {\textbf{Parity cover}\\[6pt]
Objects: fundamental cycles\\
Log number of shores: $O(q\log(2+n/q))$\\
Object price: $c_e+d_T(e)$};
\node[box] (direct) at (6.4,0) {\textbf{Direct cuts}\\[6pt]
Objects: original edges\\
Log number of shores: $O(n)$\\
Object price: $c_e$};
\node[draw=black!35,fill=black!2,rounded corners=2pt,align=center,text width=11.7cm,inner sep=8pt] at (3.2,-2.15)
{Both use the same multiplicity discount $k-q+1$,\\
the same capped growth, and the same exponential-moment bound.};
\end{tikzpicture}
\caption{The two algorithms make complementary choices. The
parity-cover algorithm compresses the family of shores using a spanning
tree, while the direct-cut algorithm avoids the cost of tree paths.
Their bounds can be compared before the first request arrives.}
\label{fig:two-algorithms}
\end{figure}

\subsection{Nonuniform requirements in one covering process}
\label{sec:overview-nonuniform}

The same approach extends to arbitrary requirements, without dividing the
requests into requirement classes. For a request of requirement $r$ and a
deficient separator of size $\ell$, the parity row has comparator mass at
least $b=r-\ell$. Instead of normalizing every row by a common slack, the
algorithm grows the fractional mass of the row toward $b$, with an
additive seed $b/m$, and the dual objective charges each update in
proportion to $b$. At the relaxed target, we have
\[
 \ell<\frac r{1+\varepsilon}
 \quad\Longrightarrow\quad b>\varepsilon\ell.
\]
Thus a forced repair at a separator of size $\ell$ requires an investment
exceeding $\varepsilon\ell$. The probability of such an event decreases
exponentially in $\ell$, while the number of possible separators of size
$\ell$ is at most $m^\ell$. Choosing the amplification
$O(\varepsilon^{-1}\log(m+1))$ makes the sum over all separator sizes
converge. This is why we need no requirement classes, no advance
knowledge of the maximum requirement, and no extra logarithm of the
maximum requirement.

\section{Capped Covering with Adaptively Exposed Rows}
\label{sec:capped}

In this section, we present the covering procedure that all our uniform
algorithms share, and we prove its guarantee once and for all. The
procedure is stated for abstract objects and rows. In the direct-cut
algorithm of \cref{sec:direct}, the objects are edges; in the
parity-cover algorithm of \cref{sec:parity}, they are fundamental cycles.

\paragraph{Objects, rows, and parameters.}
Let $\mathcal U$ be a fixed set of at most $m$ objects, where each object
$e$ has a positive price $d_e$. The algorithm maintains a set $I$ of
\emph{selected} objects, initially empty, and a nonnegative variable
$x_e$ for every object, initially zero. A coordinate $x_e$ stops changing
once its object is selected. A \emph{row} is a nonempty set
$A\subseteq\mathcal U\setminus I$ of objects that are not yet selected,
and we say that the row is \emph{hit} once one of its objects is
selected. Rows may depend on the entire history of the algorithm up to
the moment they are exposed.

We fix an integer amplification $a\ge1$ and set
\begin{equation}
 \eta=\min\{1,m/a\},\qquad
 \theta(a)=\frac54\ln\left(1+\frac{3m}{2a}\right),\qquad
 \beta=1-e^{-1}.
 \label{eq:capped-parameters}
\end{equation}
The quantity $\theta(a)$ will bound the dual load on every object. Our
purchase rule only uses coins whose probabilities are powers of two. For
$0<p\le1$, let $\dyad(p)$ be the smallest number of the form $2^{-h}$,
with $h\in\mathbb Z_{\ge0}$, that is at least $p$, and set $\dyad(0)=0$.
Thus every coin can be flipped with fair random bits.

\paragraph{The procedure.}
\Cref{alg:cover} processes one row $A$. While the row is not hit and its
fractional mass is below one, the procedure increases the coordinates of
$A$ multiplicatively, gives every object of $A$ a fresh chance to be
selected, and selects every object whose coordinate has reached the cap
$1/a$. If the mass reaches one before the row is hit, the procedure
selects a cheapest object of the row; we call this a \emph{forced
repair}.

\begin{algorithm}[t]
\caption{$\CoverRow(A)$: the shared capped purchase procedure}
\label{alg:cover}
\begin{algorithmic}[1]
\REQUIRE A nonempty row $A\subseteq\mathcal U\setminus I$; fixed prices $d_e>0$.
\WHILE{$A\cap I=\varnothing$ and $\sum_{e\in A}x_e<1$}
 \STATE $\gamma\gets(\eta/4)\min_{e\in A}d_e$.
 \STATE Compute $\Delta x_e\gets(\gamma/d_e)(x_e+1/m)$ for every $e\in A$.
 \STATE $x_e\gets x_e+\Delta x_e$ for every $e\in A$, simultaneously.
 \STATE Independently mark each $e\in A$ with probability
 $\dyad(\min\{1,a\Delta x_e\})$.
 \STATE Select every marked object and insert it into $I$.
 \STATE Select every remaining $e\in A$ with $x_e\ge1/a$, and insert it into $I$.
 \STATE Finish this entire batch before inspecting another row.
\ENDWHILE
\IF{$A\cap I=\varnothing$}
 \STATE Select a minimum-price object of $A$ and insert it into $I$.
 \COMMENT{forced repair}
\ENDIF
\end{algorithmic}
\end{algorithm}

\paragraph{Assumptions of the analysis.}
We condition on any preprocessing randomness (for instance, the sampled
spanning tree), so that all prices, shore incidences, and parameters
below are fixed. We make the following assumptions about the prefix
under analysis.
\begin{itemize}
 \item There is a fixed nonnegative vector $z$, independent of the
 purchase coins, such that $z(A)\ge1$ for every row $A$ that can be
 exposed. In our applications, $z$ is the comparator scaled by
 $1/\sigma$.
 \item There is a deterministic upper bound $\tau$ on the number of
 calls to \cref{alg:cover}.
 \item Every row is associated with a member of a fixed family
 $\mathcal S$ of at most $\nu$ shores, and each member $S$ has a fixed
 set of incident objects $\mathcal U_S\subseteq\mathcal U$. For a
 positive integer $q$, at every forced repair of a row $A$ associated
 with $S$, we have
 \begin{equation}
 A\subseteq \mathcal U_S
 \qquad\text{and}\qquad |I\cap \mathcal U_S|\le q-1.
 \label{eq:shore-hypotheses}
 \end{equation}
\end{itemize}
Notice that the hypotheses \eqref{eq:shore-hypotheses} concern only the
state of the algorithm at forced repairs. The family $\mathcal S$ can be
exponentially large, and the algorithm never needs to generate it.

\begin{lemma}[Capped adaptive covering]
\label{lem:capped}
Under the preceding assumptions, if
\begin{equation}
 \nu\exp(-\beta a+q-1)\le\frac1{2\tau},
 \label{eq:amplification-condition}
\end{equation}
then the expected total price of the objects selected by
\cref{alg:cover} is at most
\begin{equation}
 \left(6a\theta(a)+\frac12\right)d^\top z.
 \label{eq:generic-cost}
\end{equation}
Moreover, every call selects a new object, and every call performs
$O(\log(m+1))$ updates, for every $a\ge1$.
\end{lemma}

We prove \cref{lem:capped} in three steps. In \cref{sec:capped-growth},
we bound the cost of the fractional solution and of the purchases that it
triggers. In \cref{sec:capped-moment}, we show that forced repairs are
unlikely, which bounds their expected cost. Finally, in
\cref{sec:capped-termination}, we bound the number of updates in a
single call.

\subsection{Fractional growth and the cap}
\label{sec:capped-growth}

We begin with two simple consequences of the cap. A coordinate that is
updated is below $1/a$, since otherwise its object would already be
selected. Since $\eta\le1$ and $\eta\le m/a$, every update satisfies
\begin{equation}
 \Delta x_e<\frac\eta4\left(\frac1a+\frac1m\right)
 \le\frac1{2a}.
 \label{eq:cap-increment}
\end{equation}
Consequently, $x_e<3/(2a)$ at all times and $a\Delta x_e\le1/2$. A
coordinate that crosses its threshold is immediately selected and frozen.
Let $\pi_e$ be the probability of the fresh coin of $e$ in an update.
Since $\dyad$ rounds up to the next power of two, we have
\begin{equation}
 a\Delta x_e\le \pi_e\le2a\Delta x_e
 \qquad\text{and}\qquad 1-\pi_e\le\exp(-a\Delta x_e).
 \label{eq:dyadic-bounds}
\end{equation}

Next, we bound the fractional cost $d^\top x=\sum_e d_ex_e$ by a dual
argument. We assign a dual charge $\gamma_j$ to each update $j$, which is
performed on a row $A_j$. For a fixed object $e$, every update that
involves $e$ multiplies $x_e+1/m$ by $1+\gamma_j/d_e$, and therefore
\[
 1+mx_e=\prod_{j:e\in A_j}\left(1+\frac{\gamma_j}{d_e}\right).
\]
Since $0\le\gamma_j/d_e\le1/4$ and $\ln(1+t)\ge4t/5$ for
$0\le t\le1/4$, taking logarithms gives
\begin{equation}
 \sum_{j:e\in A_j}\gamma_j
 \le\frac54d_e\ln(1+mx_e)\le \theta(a)\,d_e,
 \label{eq:dual-load}
\end{equation}
where the last inequality uses $x_e<3/(2a)$. In other words, the cap
bounds the total charge on every object by $\theta(a)d_e$. On the other
hand, every update begins with $x(A_j)<1$, and therefore it increases the
fractional cost by
\[
 \Delta(d^\top x)=\gamma_j\left(x(A_j)+\frac{|A_j|}{m}\right)<2\gamma_j.
\]
Using $z(A_j)\ge1$ and summing over all updates, we obtain the pathwise
inequality
\begin{equation}
 d^\top x\le2\sum_j\gamma_j
 \le2\sum_j\gamma_j\,z(A_j)
 =2\sum_e z_e\sum_{j:e\in A_j}\gamma_j
 \le2\theta(a)\,d^\top z.
 \label{eq:generic-fractional}
\end{equation}
We remark that the rows need not become fractionally covered. An early
integral purchase may end a call before the mass of its row reaches one,
and the dual-load argument still applies.

It remains to relate the purchases to the fractional cost. Conditional on
the history just before an update, the expected price of the objects
selected by its coins is at most $2a\Delta(d^\top x)$, by
\eqref{eq:dyadic-bounds}. On the other hand, every object selected by the
threshold rule has $x_e\ge1/a$ and is selected only once, so the total
price of such objects is at most $a\,d^\top x$ on every execution. Thus
these two types of purchases have expected total price at most
$3a\E[d^\top x]\le6a\theta(a)\,d^\top z$.

\subsection{One exponential moment for each fixed shore}
\label{sec:capped-moment}

We now bound the probability of a forced repair. The difficulty is that
the rows are chosen adaptively, so we cannot fix a row in advance and
analyze it in isolation. Instead, we fix a shore $S\in\mathcal S$ and
follow its incident objects throughout the execution. We consider
checkpoints after each complete update-and-purchase batch and immediately
before each forced repair, and define
\[
 \psi_S=\exp\bigl(\beta a\,x(\mathcal U_S)-|I\cap \mathcal U_S|\bigr).
\]

We claim that $\psi_S$ is a supermartingale. Consider one batch. The
increments $\Delta x_e$ are determined before the fresh coins are
flipped, and a coordinate is updated only when its object is not
selected. For each updated incident object, its independent coin
selects it with probability $\pi_e\ge a\Delta x_e$. Let $\Delta_S$ be the
number of incident objects selected in the batch. Since
$1-\beta\pi_e\le\exp(-\beta\pi_e)$, we obtain
\[
 \E\left[e^{-\Delta_S}\mid\text{preceding history}\right]
 \le\prod_{e\in \mathcal U_S}(1-\beta \pi_e)
 \le\exp\left(-\beta a\sum_{e\in \mathcal U_S}\Delta x_e\right).
\]
Here the first inequality uses $e^{-1}=1-\beta$ and allows additional
threshold purchases, which only decrease $e^{-\Delta_S}$. The right-hand
side exactly cancels the factor by which the increase of
$x(\mathcal U_S)$ multiplies $\psi_S$. Forced purchases also only
decrease $\psi_S$. Hence $\psi_S$ is a nonnegative supermartingale, and
its initial value is at most one.

Now consider a forced repair of a row $A$ associated with $S$. At this
moment the row has mass at least one, so by \eqref{eq:shore-hypotheses}
we have $x(\mathcal U_S)\ge x(A)\ge1$ and $|I\cap\mathcal U_S|\le q-1$.
Thus $\psi_S\ge\exp(\beta a-q+1)$. By the maximal inequality for
nonnegative supermartingales and a union bound over the fixed family
$\mathcal S$,
\begin{equation}
 \Prb(\text{some forced repair occurs})
 \le \nu\exp(-\beta a+q-1)\le\frac1{2\tau}.
 \label{eq:generic-repair-prob}
\end{equation}
For completeness, we give a direct proof of the maximal inequality in our
setting. Stop the process at the first checkpoint at which it reaches the
threshold $\exp(\beta a-q+1)$, or at the last checkpoint. The stopped
value has expectation at most one, so by Markov's inequality the
threshold is reached with probability at most its reciprocal. Since the
number of updates is bounded deterministically (see
\cref{sec:capped-termination}), no limiting stopping argument is
needed.

Finally, we bound the cost of forced repairs. Since $z(A)\ge1$, the
cheapest object of any exposed row costs at most
$\min_{e\in A}d_e\cdot z(A)\le d^\top z$. At most $\tau$ forced repairs
occur, so their total price is at most $\tau\,d^\top z$ on every
execution, and it is zero unless the event in
\eqref{eq:generic-repair-prob} occurs. Hence the expected price of
forced repairs is at most $d^\top z/2$. Together with the bound of
\cref{sec:capped-growth}, this proves the cost bound
\eqref{eq:generic-cost}.

\subsection{Termination of a single call}
\label{sec:capped-termination}

Finally, we bound the number of updates in one call of
\cref{alg:cover} on a row $A$. Fix a cheapest object $e_0$ of $A$. For
$e_0$, the step size satisfies $\gamma/d_{e_0}=\eta/4$, and therefore
every update multiplies $x_{e_0}+1/m$ by $1+\eta/4$. Starting from zero,
$x_{e_0}$ reaches the threshold $1/a$ within
\begin{equation}
 \ceil{\frac{\ln(1+m/a)}{\ln(1+\eta/4)}}
 \label{eq:row-updates}
\end{equation}
updates, unless another purchase or a forced repair ends the call
earlier. If $a\le m$, then $\eta=1$, and the bound is $O(\log(m+1))$. If
$a>m$, then $\eta=m/a<1$, and the inequality
$(1+\eta/4)^4\ge1+\eta$ shows that at most four updates suffice. In both
cases, the call selects a new object: either the threshold rule, a coin,
or a forced repair hits the row. This completes the proof of
\cref{lem:capped}.

\section{The Direct-Cut Algorithm}
\label{sec:direct}

We begin with the simpler of our two uniform algorithms, which applies
\cref{lem:capped} directly to the cuts of the bought graph. This
algorithm does not use a spanning tree. Its objects are the original
edges of positive cost, and the price of an edge is its cost, that is,
$d_e=c_e$. All zero-cost edges are bought at the beginning.

Whenever the current request $(u,v)$ has connectivity $\ell<q$ in the
bought graph $H$, the algorithm finds a minimum-cut shore $S$ for $u$ and
$v$ in $H$ and exposes the row
\begin{equation}
 A=\delta_G(S)\setminus H
 \label{eq:direct-row}
\end{equation}
of unbought edges that cross $S$. It then runs \cref{alg:cover} on this
row with the amplification $a_0$ that we specify in
\cref{sec:parameters}, and selecting an object means buying the
corresponding edge. Since every edge of $A$ crosses $S$, buying it
increases the size of this cut. After each call, the algorithm recomputes
a minimum cut, and it continues until the request is served.
\Cref{alg:direct} summarizes the algorithm.

\begin{algorithm}[t]
\caption{Uniform direct cuts}
\label{alg:direct}
\begin{algorithmic}[1]
\STATE Buy all zero-cost edges; initialize $I\gets\varnothing$ and $x_e\gets0$ on positive-cost edges.
\STATE Set $d_e\gets c_e$, $a\gets a_0$, and $\eta\gets\min\{1,m/a_0\}$.
\FOR{each arriving pair $(u,v)$}
 \WHILE{$\lambda_H(u,v)<q$}
  \STATE Compute a minimum-cut shore $S$ for $u$ and $v$ in $H$.
  \STATE Set $A\gets\delta_G(S)\setminus H$.
  \STATE Call $\CoverRow(A)$; selecting $e$ buys that original edge.
 \ENDWHILE
\ENDFOR
\end{algorithmic}
\end{algorithm}

To analyze the algorithm, we verify the three assumptions of
\cref{lem:capped}: a fractional cover $z$ of all rows, a bound $\tau_0$ on
the number of calls, and a family of shores that satisfies
\eqref{eq:shore-hypotheses}.

\paragraph{The comparator covers every row.}
Fix any feasible full-$k$ cut vector $y$. Since $S$ separates a requested
pair, we have $y(\delta_G(S))\ge k$, and since $y_e\le1$ for every edge,
\[
 y(A)\ge y(\delta_G(S))-|\delta_H(S)|\ge k-\ell\ge \sigma.
\]
Thus $z=y/\sigma$ covers every exposed row.

\paragraph{The number of calls.}
To bound the number of calls, we use the connectivity classes of the
bought graph. For an integer $h\ge1$ and an undirected graph $J$, the
relation defined by $u\sim_h v$ if $u=v$ or $\lambda_J(u,v)\ge h$ is an
equivalence relation on the vertices. Transitivity holds because every
cut that separates $u$ and $w$ separates at least one of the pairs $u,v$
and $v,w$. Let $\kappa_h(J)$ be the number of its equivalence classes.
Adding edges to $J$ can only merge classes, and therefore
$\kappa_h(J)$ can only decrease.

\begin{lemma}[Connectivity-rank bound for direct cuts]
\label{lem:rank-direct}
The total number of calls to \cref{alg:cover} in \cref{alg:direct} is
at most
\begin{equation}
 \tau_0=\min\{m,q(n-1)\}.
 \label{eq:direct-rank}
\end{equation}
\end{lemma}

\begin{proof}
We use the integer potential
\[
 \phi_0(H)=\sum_{h=1}^q(n-\kappa_h(H)),
\]
which is nondecreasing and at most $q(n-1)$. Consider the first edge $e$
selected in a call, and let $u'$ and $v'$ be its endpoints. Since $e$
crosses $S$, we have $h:=\lambda_H(u',v')\le\ell<q$. Adding $e$
increases the connectivity between $u'$ and $v'$ to $h+1$, and therefore
merges the $(h+1)$-connectivity classes of $u'$ and $v'$. Hence each call
increases $\phi_0$ by at least one, which gives the bound $q(n-1)$. The
bound $m$ holds because each call selects a new edge.
\end{proof}

Since the potential includes the level $h=1$, the lemma also applies when
the bought graph is disconnected at the beginning.

\paragraph{The family of shores.}
We represent every nontrivial shore by its side that does not contain a
fixed reference vertex. Thus there are fewer than $\nu_0=2^{n-1}$ such
shores. For each shore $S$, we let $\mathcal U_S$ be the set of
positive-cost edges in $\delta_G(S)$. At a forced repair, the row
satisfies $A\subseteq \mathcal U_S$, and at most $|\delta_H(S)|\le q-1$
selected objects cross $S$. Thus the hypotheses
\eqref{eq:shore-hypotheses} hold.

\paragraph{The cost.}
All assumptions of \cref{lem:capped} are satisfied. Hence, if the
amplification $a_0$ satisfies \eqref{eq:amplification-condition} with
$\nu=\nu_0$ and $\tau=\tau_0$, then
\begin{equation}
 \E[c(H_i)]\le
 \frac{6a_0\theta(a_0)+1/2}{\sigma}\,c^\top y.
 \label{eq:direct-exact}
\end{equation}
Notice that this algorithm pays nothing for tree paths. In
\cref{sec:parameters}, we show that the choice $a_0=\Theta(n+q)$
satisfies \eqref{eq:amplification-condition}, which proves
\eqref{eq:direct-tradeoff}.

\section{The Parity-Cover Algorithm}
\label{sec:parity}

In this section, we present the parity-cover algorithm and prove the
parity bound \eqref{eq:parity-tradeoff} of \cref{thm:tradeoff}, up to the
choice of parameters in \cref{sec:parameters}. The algorithm samples one
spanning tree $T$ satisfying \eqref{eq:stretch} at the beginning and fixes
a reference vertex $v_0$ as its root. Until we average over the tree at
the end of this section, we condition on $T$.

\paragraph{Objects and prices.}
The objects of the parity-cover algorithm are the chords $e\notin T$.
Selecting a chord $e$ means buying every missing edge of its fundamental
cycle $C_e$, and we charge the full price
\[
 d_e=c_e+d_T(e)
\]
for it, even when a part of the cycle is already bought. Each selected
chord is recorded once in $I$. In addition, the algorithm buys the tree
path of every request, and we charge these paths separately by the cost of
their union. Since the bought graph is always the union of these
purchases, the charges are an upper bound on the amount that the
algorithm actually pays. We also write
$d^\top y=\sum_{e\notin T}d_ey_e$ for a vector $y$; this is the price of
the comparator on the chords.

\paragraph{Zero costs.}
All zero-cost edges are bought at the beginning. To make sure that every
zero-cost chord also has a zero price, we choose the tree as follows. We
choose a zero-cost spanning forest inside each component of zero-cost
edges, sample a tree after contracting these components \emph{for the
purpose of tree selection only}, and lift the sampled tree using the
forest. Then every zero-cost edge has a zero-cost tree path. All
zero-price chords are inserted into $I$ at the beginning; their cycles
are already bought. A violated row never contains a selected object, and
therefore all denominators in the updates of \cref{alg:cover} are
positive. We emphasize that the original vertices and all unit-capacity
edges are kept for the connectivity computations: contracting the
zero-cost components in those computations would be incorrect.

\subsection{Parity shores}

For any set of edges $Q$, we label every vertex by the parity of the
number of edges of $Q$ on its tree path from the root, and we define
\begin{equation}
 \chi_Q(v)=|P_T(v_0,v)\cap Q|\pmod2,
 \qquad S_Q=\{v:\chi_Q(v)=1\},
 \qquad A_Q=\delta_G(S_Q)\setminus Q.
 \label{eq:parity-row}
\end{equation}
The labels of the two endpoints of a tree edge differ exactly when the
edge belongs to $Q$. Thus, for every $Q$,
\begin{equation}
 \delta_T(S_Q)=Q\cap T.
 \label{eq:tree-boundary}
\end{equation}
The following lemma collects the properties of the parity row $A_Q$ that
we need.

\begin{lemma}[Parity and multiplicity]
\label{lem:parity}
Suppose that $H$ contains $P_T(u,v)$ and every cycle $C_e$ with $e\in I$.
Let $Q$ be a minimum $u$--$v$ separator of $H$, of size $\ell$. Then the
following hold.
\begin{enumerate}
 \item $S_Q$ separates $u$ and $v$, all members of $A_Q$ are chords, and
 $A_Q\cap I=\varnothing$.
 \item For every $e\in A_Q$, adding $C_e$ to $H$ destroys $Q$ as a
 separator.
 \item Every feasible full-$k$ cut vector $y$ for this request satisfies
 \begin{equation}
 y(A_Q)\ge k-\ell.
 \label{eq:parity-comparator}
 \end{equation}
\end{enumerate}
\end{lemma}

\begin{proof}
Let $U$ be the set of vertices reachable from $u$ in $H-Q$. Then
$\delta_H(U)\subseteq Q$, and since $Q$ is a minimum separator, equality
holds. The bought tree path from $u$ to $v$ crosses this cut an odd
number of times, and therefore $\chi_Q(u)\ne\chi_Q(v)$; that is, $S_Q$
separates $u$ and $v$. By \eqref{eq:tree-boundary}, a tree edge outside
$Q$ has equal labels at its endpoints, so $A_Q$ contains only chords.

Now let $e\in A_Q$, with endpoints $u'$ and $v'$. Since the labels of
$u'$ and $v'$ differ, its fundamental cycle satisfies
\[
 |C_e\cap Q|\equiv|P_T(u',v')\cap Q|
 \equiv\chi_Q(u')+\chi_Q(v')\equiv1\pmod2.
\]
Suppose, for the sake of contradiction, that $Q$ still separated $u$ and
$v$ in $J=H\cup C_e$, and let $U'$ be the set of vertices reachable from
$u$ in $J-Q$. Then
\[
 \ell\le|\delta_H(U')|\le|\delta_J(U')|\le|Q|=\ell.
\]
It follows that $\delta_J(U')=Q$, which contradicts the fact that a cycle
and a cut always intersect in an even number of edges. This proves the
second claim. In particular, a cycle that is already contained in $H$
cannot have its chord in $A_Q$, and hence $A_Q\cap I=\varnothing$.

Finally, $S_Q$ separates the request, so $y(\delta_G(S_Q))\ge k$.
Removing $Q$ loses at most $|Q|=\ell$ units, since $y_e\le1$ for every
edge. This proves \eqref{eq:parity-comparator}.
\end{proof}

\subsection{The algorithm}

The parity-cover algorithm is given in \cref{alg:parity}. On each
request $(u,v)$, it first buys the tree path $P_T(u,v)$. Then, as long as
the connectivity of $u$ and $v$ is below $q$, it computes a minimum
separator $Q$, forms the parity row $A_Q$ by a traversal of $T$ and a
scan of the edges, and runs \cref{alg:cover} on it. The algorithm
recomputes a minimum cut after each call, since repairing one separator
need not increase the minimum-cut value immediately. For $q=1$, the
algorithm only buys requested tree paths.

\begin{algorithm}[t]
\caption{Uniform parity cover for a common full requirement $k$ and target $q$}
\label{alg:parity}
\begin{algorithmic}[1]
\STATE Sample $T\subseteq G$ satisfying \eqref{eq:stretch}; choose a reference root $v_0$.
\STATE Buy zero-cost edges; put their zero-price chords into $I$.
\STATE For every other chord $e$, set $d_e\gets c_e+d_T(e)$ and $x_e\gets0$.
\STATE If $q\ge2$, set $a\gets a_T$ and $\eta\gets\min\{1,m/a\}$ from \cref{sec:parameters}.
\FOR{each arriving pair $(u,v)$}
 \STATE Buy every missing edge of $P_T(u,v)$.
 \WHILE{$\lambda_H(u,v)<q$}
  \STATE Compute a minimum $u$--$v$ separator $Q$ in $H$.
  \STATE Traverse $T$ to compute $\chi_Q$ and $S_Q$ in \eqref{eq:parity-row}.
  \STATE Form $A_Q$ by scanning $E$; call $\CoverRow(A_Q)$.
  \STATE A selected object $e$ buys the whole cycle $C_e$.
 \ENDWHILE
\ENDFOR
\end{algorithmic}
\end{algorithm}

\subsection{A bound on the number of calls}

We now bound the number of calls to \cref{alg:cover}. As in
\cref{lem:rank-direct}, we use the connectivity classes $\kappa_h$
defined in \cref{sec:direct}. This time, however, we apply them to a
virtual graph that consists of the tree and the selected chords.

\begin{lemma}[Connectivity-rank bound]
\label{lem:rank}
For $q\ge2$, the total number of calls to \cref{alg:cover} in
\cref{alg:parity} is at most
\begin{equation}
 \tau_T=\min\{m,(q-1)(n-1)\}.
 \label{eq:parity-rank}
\end{equation}
\end{lemma}

\begin{proof}
Consider the virtual graph $K=T\cup I$ and the integer potential
\[
 \phi_T(K)=\sum_{h=2}^q(n-\kappa_h(K)),
\]
which is nondecreasing and at most $(q-1)(n-1)$. Consider a call on the
row of a separator $Q$, and let $e$ be the first new chord selected in
this call, with endpoints $u'$ and $v'$. Immediately before $e$ is
selected, we have
\[
 \delta_K(S_Q)\subseteq Q.
\]
Indeed, \eqref{eq:tree-boundary} accounts for all tree edges that cross
$S_Q$, and \cref{lem:parity} forces all previously selected chords that
cross $S_Q$ into $Q$. Since $e\in A_Q$ crosses $S_Q$, it follows that
$1\le h:=\lambda_K(u',v')\le|Q|<q$, where the lower bound holds because
$K$ contains the spanning tree $T$. Adding the edge $e$ increases every
$u'$--$v'$ cut by one, so the new connectivity is exactly $h+1$. Thus the
$(h+1)$-connectivity classes that contain $u'$ and $v'$ merge, and
$\phi_T$ increases by at least one. Further purchases cannot decrease
$\phi_T$. This gives the bound $(q-1)(n-1)$, and the bound $m$ holds
because each call selects a distinct chord.
\end{proof}

We remark that the virtual graph is used only in this proof. Its tree
edges are paid for when requested paths or selected cycles use them; we
never assume any free physical capacity.

\subsection{Applying capped covering}

We now fix a prefix, its comparator $y$, and the tree $T$. We set
\begin{equation}
 z_e=\frac{y_e}{\sigma}\qquad\text{for every chord } e\notin T.
 \label{eq:tree-comparator}
\end{equation}
By \cref{lem:parity}, every exposed row $A_Q$ satisfies
$y(A_Q)\ge k-\ell\ge k-q+1=\sigma$, and hence $z$ covers every row.

Next, we define the family of shores. A parity shore is determined by the
set $Q\cap T$, which is a nonempty set of at most $q-1$ tree edges.
Therefore, the family of parity shores has size at most
\begin{equation}
 \nu_T=\sum_{j=1}^{\min\{q-1,n-1\}}\binom{n-1}{j}.
 \label{eq:parity-entropy}
\end{equation}
We restrict the set of objects and the vector $z$ to chords of positive
price; the zero-price chords are selected at the beginning and never
occur in a row. For a shore $S$, we let $\mathcal U_S$ be the set of
positive-price chords in $\delta_G(S)\setminus T$. At a forced repair of
the row of $Q$, the row is contained in $\mathcal U_{S_Q}$, and all
selected chords that cross $S_Q$ lie in $Q$. More precisely,
\[
 |I\cap \mathcal U_{S_Q}|\le |Q|-|Q\cap T|\le q-2,
\]
because the requested tree path forces $|Q\cap T|\ge1$. The weaker bound
$q-1$ suffices for \cref{lem:capped}. Hence, if the amplification
satisfies \eqref{eq:amplification-condition} with $\nu=\nu_T$ and
$\tau=\tau_T$, the conditional expected price of the selected cycles is
at most
\begin{equation}
 \frac{6a\theta(a)+1/2}{\sigma}\,d^\top y.
 \label{eq:parity-cycle-cost}
\end{equation}

\paragraph{The cost of tree paths.}
It remains to bound the cost of the requested tree paths. Let
$\mathcal B_i$ be the union of the requested tree paths in the prefix,
excluding tree edges that are bought only as parts of cycles. For each
tree edge $f$, let $S_f$ be the vertex set of one of the two components
of $T-f$. An edge $e$ crosses $S_f$ exactly when $f\in P_T(e)$, and
therefore double counting gives
\begin{equation}
 \sum_{e\in E}y_e d_T(e)
 =\sum_{f\in T}c_f\, y(\delta_G(S_f))\ge k\,c(\mathcal B_i),
 \label{eq:backbone}
\end{equation}
where the inequality holds because every $f\in\mathcal B_i$ lies on the
tree path of a requested pair, so $S_f$ separates that pair and
$y(\delta_G(S_f))\ge k$.

\paragraph{The total cost.}
We now average over the tree. Since $y$ is fixed independently of $T$,
the stretch guarantee \eqref{eq:stretch} gives
$\E[\sum_e y_ed_T(e)]\le \rho\, c^\top y$ and
$\E[d^\top y]\le(1+\rho)\,c^\top y$. Combining these bounds with
\eqref{eq:parity-cycle-cost} and \eqref{eq:backbone}, we obtain
\begin{equation}
 \E[c(H_i)]\le
 \left[\frac{\rho}k+
 \frac{1+\rho}\sigma\left(6a\theta(a)+\frac12\right)\right]c^\top y.
 \label{eq:parity-exact}
\end{equation}
Choosing an optimal $y$ proves the parity bound in terms of the
parameters; we simplify it in \cref{sec:parameters}.

Finally, the algorithm is always feasible. Every deficient request
supplies a nonempty row by \cref{lem:parity}, each call selects a new
chord by \cref{lem:capped}, and by \cref{lem:rank} the process cannot
continue indefinitely.

\section{Arbitrary Requirements without Requirement Classes}
\label{sec:nonuniform}

In this section, we prove \cref{thm:nonuniform}. We keep the parity
construction of \cref{sec:parity}, including the sampled tree, the cycle
prices $d_e=c_e+d_T(e)$, and the treatment of zero costs. The difference
is in the covering procedure. Different requests now have different
requirements, and therefore there is no common slack by which we could
normalize the rows. Instead, we replace the normalized process of
\cref{alg:cover} by a fractional multicover process, in which every row
has its own threshold.

\paragraph{The algorithm.}
Fix $\varepsilon$, and let $a_*$ be the amplification that we choose in
\cref{sec:parameters}. We initialize $x_e=0$ for every chord, and we
initialize $I$ with all zero-price chords, whose cycles are bought for
free at the beginning. As before, the coordinates of selected chords are
frozen. On a request $(u,v,r)$, the algorithm first buys the tree path
$P_T(u,v)$ and sets the target $q=\ceil{r/(1+\varepsilon)}$.

As long as a minimum separator $Q$ has size $\ell<q$, the algorithm forms
the parity row $A_Q$ and sets its threshold to $b=r-\ell$. By
\cref{lem:parity}, applied with $r$ in place of $k$, every fixed
full-requirement comparator $y$ satisfies
\begin{equation}
 y(A_Q)\ge b,\qquad
 b>\varepsilon\ell,\qquad\ell\ge1.
 \label{eq:nonuniform-density}
\end{equation}
The middle inequality follows from $\ell\le q-1<r/(1+\varepsilon)$, and
the last one holds because $H$ contains the tree path $P_T(u,v)$. The
algorithm then runs \cref{alg:multicover} on the row $A_Q$ with threshold
$b$. After each call, it recomputes a minimum cut and continues until the
current target is met.

\begin{algorithm}[t]
\caption{$\MultiCoverRow(A,b)$ for a nonuniform request}
\label{alg:multicover}
\begin{algorithmic}[1]
\WHILE{$A\cap I=\varnothing$ and $x(A)<b$}
 \STATE $\gamma\gets\frac14\min_{e\in A}d_e$.
 \STATE Compute $\Delta x_e\gets(\gamma/d_e)(x_e+b/m)$ for every $e\in A$.
 \STATE Increase all $x_e$ simultaneously by $\Delta x_e$.
 \STATE Independently select each $e\in A$ with probability
 $\dyad(\min\{1,a_*\Delta x_e\})$.
 \STATE Buy every selected cycle, insert its chord into $I$, and finish the batch.
\ENDWHILE
\IF{$A\cap I=\varnothing$}
 \STATE Buy a cheapest cycle of $A$ and insert its chord into $I$.
\ENDIF
\end{algorithmic}
\end{algorithm}

\Cref{alg:multicover} differs from \cref{alg:cover} in two ways. First,
the additive seed and the threshold of a row are proportional to its
residual requirement $b$. Second, there is no deterministic cap at $1/a_*$
in this version: the varying row threshold already provides the growth
bound that we need. Since each call selects a new chord, at most $m$
calls occur.

\subsection{One comparator for all residual thresholds}

We now bound the cost of the fractional solution and of the sampled
purchases. Fix a prefix and its feasible cut vector $y$. Let
$k=\max_{j\le i}r_j\le m$ be the maximum requirement in the prefix; this
is a parameter of the analysis, not an input of the algorithm. As in
\cref{sec:parity}, we write $d^\top y=\sum_{e\notin T}d_ey_e$ and
$d^\top x=\sum_{e\notin T}d_ex_e$, and we set
\[
 \theta_* =\frac54\ln\left(1+\frac32mk\right).
\]

We first bound the dual load on every chord. An updated coordinate has
value below $b\le k$ before the update, and since $\gamma/d_e\le1/4$, it
always remains at most $3k/2$. Moreover, an update multiplies the
quantity $1+mx_e$ by at least $1+\gamma/d_e$: writing $x_e^{\rm new}$
for the updated value, we have
\[
 1+mx_e^{\rm new}
 =\left(1+\frac\gamma{d_e}\right)(1+mx_e)+\frac\gamma{d_e}(b-1)
 \ge\left(1+\frac\gamma{d_e}\right)(1+mx_e).
\]
The same logarithmic calculation as in \eqref{eq:dual-load} therefore
gives
\[
 \sum_{j:e\in A_j}\gamma_j\le \theta_*\,d_e.
\]
On the other hand, an update $j$ on a row $A_j$ with threshold $b_j$
increases the fractional cost by less than $2b_j\gamma_j$, since
$x(A_j)<b_j$ and the seed contributes $|A_j|b_j/m\le b_j$. Using
$y(A_j)\ge b_j$, we obtain
\begin{equation}
 d^\top x\le2\sum_jb_j\gamma_j
 \le2\sum_e y_e\sum_{j:e\in A_j}\gamma_j
 \le2\theta_*\,d^\top y.
 \label{eq:nonuniform-frac}
\end{equation}
Notice that a single comparator pays for all rows, regardless of their
thresholds. The dyadic coin rule satisfies $\pi_e\le2a_*\Delta x_e$ and
$1-\pi_e\le e^{-a_*\Delta x_e}$, including when $a_*\Delta x_e\ge1$.
Hence the expected price of the sampled purchases is at most
$4a_*\theta_*\,d^\top y$.

\subsection{A repair bound stratified by separator size}

It remains to bound the cost of forced repairs. We follow the approach of
\cref{sec:capped-moment}, but this time we work directly with formal
separators. Fix a set of edges $Q$ of size $\ell\ge1$ and its row $A_Q$,
whether or not this row is ever exposed. We claim that the process
\[
 \psi_Q=\ind[I\cap A_Q=\varnothing]\exp\bigl(a_*\, x(A_Q)\bigr)
\]
is a nonnegative supermartingale. Indeed, as long as the row is not hit,
the fresh coins of an update leave it unhit with probability at most
$\exp(-a_*\sum_{e\in A_Q}\Delta x_e)$, which cancels the fractional
growth; and once an object of the row is selected, the process becomes
zero forever.

A forced repair of this row occurs only after its mass reaches
$b>\varepsilon\ell$, while the row is still unhit. By the same maximal
inequality as in \cref{sec:capped-moment}, this happens with probability
at most $e^{-a_*\varepsilon\ell}$. There are at most $m^\ell$ sets of
edges of size $\ell$. Therefore, if $a_*\varepsilon\ge2\ln(4m)$, then
\begin{equation}
 \Prb(\text{some forced repair occurs}\mid T)
 \le\sum_{\ell=1}^{m}m^\ell e^{-a_*\varepsilon\ell}
 \le\sum_{\ell=1}^{\infty}(16m)^{-\ell}
 =\frac1{16m-1}\le\frac1{2m}.
 \label{eq:nonuniform-repairs}
\end{equation}
We emphasize that this sum is only a probability estimate over a fixed
implicit family; the algorithm never enumerates these rows. The
exponential decay in $\ell$, which comes from the inequality
$b>\varepsilon\ell$, is what makes the sum over all separator sizes
converge.

By \eqref{eq:nonuniform-density}, a cheapest cycle of a row costs at most
$d^\top y/b\le d^\top y$. There are at most $m$ forced repairs, so their
expected price is at most $d^\top y/2$. Altogether, the conditional
expected price of all cycles is at most
\begin{equation}
 \left(4a_*\theta_*+\frac12\right)d^\top y.
 \label{eq:nonuniform-cycle-cost}
\end{equation}

\paragraph{The total cost.}
Every tree edge in the union of requested paths defines a cut that
separates a requested pair, and hence the comparator crosses it with
capacity at least one. As in \eqref{eq:backbone}, the union of the
requested paths therefore costs at most $\sum_e y_ed_T(e)$. Averaging
over the tree, we obtain
\begin{equation}
 \E[c(H_i)]\le
 \bigl[\rho+(1+\rho)(4a_*\theta_*+1/2)\bigr]c^\top y.
 \label{eq:nonuniform-exact}
\end{equation}
The parameter estimates in \cref{sec:parameters} complete the proof of
\cref{thm:nonuniform}.

\section{Implementation and Running Time}
\label{sec:runtime}

In this section, we show that all our algorithms run in polynomial time,
regardless of the value of $k$. Suppose that $i$ requests have been
processed. All algorithms use ordinary minimum-cut computations in the
currently bought unit-capacity graph. We fix a deterministic tie-breaking
rule for cuts and for objects of minimum price; the particular choices do
not affect the analysis. Computing a parity row takes one traversal of
$T$ and one scan of $E$, and computing a row from an actual cut takes one
scan of $E$. The cycle prices are computed once, at the beginning; the
algorithms never change these prices to reflect earlier purchases.

\paragraph{Uniform algorithms.}
Let $\MC(n,m)$ be the running time of any polynomial-time minimum-cut
routine for one pair. Let $\tau=\tau_T$ for the parity-cover algorithm
and $\tau=\tau_0$ for the direct-cut algorithm. The uniform algorithms
make $O(i+\tau)$ minimum-cut calls. By \cref{lem:capped}, they perform
$O(\tau\log(m+1))$ fractional updates, each of which scans at most $m$
objects. Expanding the paths of all selected cycles takes $O(mn)$
additional work with a direct implementation, and the requested tree
paths take $O(in)$ work. Altogether, after the preprocessing of the tree
where applicable, the algorithms perform
\begin{equation}
 O\bigl((i+\tau)\MC(n,m)
       +m\tau\log(m+1)+mn+in\bigr)
 \label{eq:runtime}
\end{equation}
arithmetic and graph operations. The direct-cut algorithm does not need
the path-expansion work. Sorting the parallel bundles for
\cref{lem:pruning} takes $O(m\log(m+1))$ comparisons.

\paragraph{The nonuniform algorithm.}
In a call of \cref{alg:multicover}, a cheapest member $e_0$ of the row
satisfies
\[
 x_{e_0}^{\rm new}+b/m=\frac54\left(x_{e_0}+b/m\right).
\]
Hence it reaches mass $b$ within $\ceil{\log_{5/4}(m+1)}$ updates,
unless a purchase ends the call earlier. Thus the nonuniform algorithm
satisfies the same bound \eqref{eq:runtime} with $\tau=m$. In
particular, no loop of our algorithms has a length proportional to
$1/\varepsilon$ or to an unknown maximum requirement.

\paragraph{Exact arithmetic.}
Our algorithms do not need to sample from real-valued distributions. A
coin of probability $2^{-h}$ is implemented by testing whether $h$ fair
random bits are all zero. The integer amplification parameters of
\cref{sec:parameters} are obtained by binomial recurrences, integer
powers, and comparisons. All costs and updates are rational: each update
multiplies or adds rational quantities that are determined by the input
prices and the displayed parameters, and over the polynomially many
updates, their representations remain polynomial in the input length.
Consequently, the algorithms run in polynomial time with exact rational
arithmetic and fair random bits, in addition to satisfying the operation
bounds above.

\section{Choosing the Parameters}
\label{sec:parameters}

In this section, we set the amplification parameters of our algorithms,
verify the inequalities used in their analyses, and complete the proofs
of \cref{thm:headline,thm:tradeoff,thm:nonuniform}. Natural logarithms
are denoted by $\ln$; changing the base of a logarithm in an asymptotic
expression affects only constants. If $m=1$, the only edge is bought
when it is requested, so we assume $m\ge2$.

\subsection{The parity-cover algorithm}

For $q\ge2$, let $\tau_T$ and $\nu_T$ be given by \eqref{eq:parity-rank}
and \eqref{eq:parity-entropy}, and set
\begin{equation}
 g_T=\ceil{\log_2(2\tau_T\nu_T)},\qquad
 a_T=2(q+g_T),\qquad
 \eta_T=\min\{1,m/a_T\}.
 \label{eq:parity-constants}
\end{equation}
Since $\beta>1/2$, we have
\[
 \beta a_T-q+1>g_T\ge\ln(2\tau_T\nu_T),
\]
which verifies \eqref{eq:amplification-condition}. Notice also that
$a_T\ge2q$.

Next, we estimate $a_T$. Let $\bar q=\min\{q-1,n-1\}$. If
$\bar q\le(n-1)/2$, the binomial estimate
$\sum_{j=0}^{\bar q}\binom{n-1}{j}\le(e(n-1)/\bar q)^{\bar q}$ bounds
$\nu_T$, and otherwise we use $\nu_T<2^{n-1}$. Together, these bounds
imply
\[
 \ln \nu_T=O\!\left(q\log\left(2+\frac nq\right)\right).
\]
Furthermore, $\tau_T\le(q-1)(n-1)$, and hence
$\log(2\tau_T)=O(\log(q+1)+\log(n+1))$. Both terms are
$O(q\log(2+n/q))$: the first is $O(q)$, and the second follows because
the function $\xi\mapsto \xi\log(2+n/\xi)$ is increasing on $\xi>0$.
Consequently,
\begin{equation}
 a_T=O\!\left(q\log\left(2+\frac nq\right)\right)
 \qquad\text{and}\qquad
 \theta(a_T)=O\!\left(\log\left(1+\frac mq\right)\right).
 \label{eq:parity-parameter-estimates}
\end{equation}
The additive $1/2$ in \eqref{eq:parity-exact} is absorbed by
$a_T\theta(a_T)$: here $a_T\le6m$, and $a_T\ln(1+3m/(2a_T))$ is bounded
below by a positive constant. Substituting these estimates in
\eqref{eq:parity-exact} and using $q\le2(q-1)$ proves
\eqref{eq:parity-tradeoff}.

For the running time, the bounds $\nu_T<2^{n-1}$, $\tau_T\le m$, $q\le m$,
and $n-1\le m$ imply
\[
 a_T\le2\bigl(q+n-1+\ceil{\log_2(2m)}\bigr)\le6m.
\]
Thus $\eta_T\ge1/6$, as required for \eqref{eq:row-updates}. The binomial
sums have integer representations of polynomial size, and we compute them
without enumerating subsets.

\subsection{The direct-cut algorithm}

For $1\le q\le k$, we set
\begin{equation}
 \nu_0=2^{n-1},\quad \tau_0=\min\{m,q(n-1)\},\quad
 g_0=\ceil{\log_2(2\tau_0\nu_0)},\quad a_0=2(q+g_0).
 \label{eq:direct-constants}
\end{equation}
Again $\beta a_0-q+1>g_0\ge\ln(2\tau_0\nu_0)$, which proves
\eqref{eq:amplification-condition}. The identity
$g_0=n-1+\ceil{\log_2(2\tau_0)}$ and the bound $\tau_0\le q(n-1)$ give
\[
 a_0=\Theta(n+q),\qquad a_0\le6m,\qquad\eta_0\ge1/6.
\]
Substituting these estimates in \eqref{eq:direct-exact} proves
\eqref{eq:direct-tradeoff}. This completes the proof of
\cref{thm:tradeoff}.

\subsection{Removing excess parallel edges}

Before we turn to the near-exact target, we show that a known common
requirement allows us to discard most parallel edges. This reduction
replaces $m$ by a quantity that is polynomial in $n$, which turns the
logarithms of $m$ in our bounds into logarithms of $n$.

\begin{lemma}[Parallel-bundle reduction]
\label{lem:pruning}
For a known common requirement $k$, retain the $k$ cheapest edges of
each parallel bundle, or the entire bundle when it has fewer than $k$
edges. This preserves $\LP_k(i)$ and $\OPT_k(i)$ for every prefix of
requests. The retained graph has at most $m'\le k\binom n2$ edges.
\end{lemma}

\begin{proof}
Clearly, a solution on the retained edges is also a solution on the
original graph. For the other direction, let $y$ be a feasible full-$k$
vector, and consider a bundle $B$. We replace the mass of $y$ on $B$ by
a total mass of $\min\{k,y(B)\}$, which we assign to the retained edges
of $B$ in nondecreasing order of cost, with at most one unit per edge.
This does not increase the cost. Now consider a required cut. A cut
contains either an entire bundle or none of it. If every bundle that
crosses the cut had mass less than $k$, the total mass of the cut is
unchanged. Otherwise, some crossing bundle had mass at least $k$, and
this bundle alone still provides $k$ units. Thus every required cut
remains feasible after all replacements. Moreover, the replacement
preserves integrality, since integral bundle masses are assigned
integrally. This proves that both optimum values are preserved, and the
bound on the number of edges is immediate.
\end{proof}

The reduction uses the known common requirement. We therefore do not
apply it in the nonuniform algorithm, which does not know the future
maximum requirement.

\subsection{The near-exact target and the better of the two algorithms}

We now consider the target $q=\ceil{k/(1+\varepsilon)}$. The ceiling
gives
\begin{equation}
 q-1<\frac{k}{1+\varepsilon},\qquad
 \sigma=k-q+1>\frac{\varepsilon k}{1+\varepsilon},\qquad
 \frac{q-1}\sigma<\frac1\varepsilon.
 \label{eq:slack-inequalities}
\end{equation}
In addition, for every $q\ge1$, we have $k/2\le q\le k$ and
$q/\sigma\le2/\varepsilon$. Therefore, the parity guarantee becomes
\begin{equation}
 O\!\left(\rho\left[\frac1k+
 \frac1\varepsilon\log\left(2+\frac nk\right)
 \log\left(1+\frac mk\right)\right]\right)\LP_k(i),
 \label{eq:near-parity}
\end{equation}
and the direct guarantee becomes
\begin{equation}
 O\!\left(\frac1\varepsilon\left(1+\frac nk\right)
 \log\left(1+\frac m{n+k}\right)\right)\LP_k(i).
 \label{eq:near-direct}
\end{equation}
For $q=1$, the guarantee of the tree paths is already at least as strong
as the asymptotic parity bound.

We apply \cref{lem:pruning} before choosing an algorithm. Since
$m'\le k\binom n2$, both logarithms involving $m$ in
\cref{eq:near-parity,eq:near-direct} become $O(\log(n+1))$, and the
term $1/k$ of the tree paths in \eqref{eq:near-parity} is absorbed. Choosing the
algorithm with the smaller of these two known upper bounds and
substituting $\rho=O(\log(n+1))$ from~\cite{Hajiaghayi26} proves
\cref{thm:headline}.

More concretely, the choice is made as follows. We use a declared
rational upper bound $\hat\rho=O(\log(n+1))$ on the stretch of the
sampler. For $q\ge2$, we replace $\theta(a)$ in
\cref{eq:parity-exact,eq:direct-exact} by the rational bound
\[
 \theta^+(a)=\frac54\ceil{\log_2\left(1+\frac{3m'}{2a}\right)},
\]
and we compare the two resulting coefficients, using $\hat\rho$ in place
of $\rho$. For the chosen parameters, these replacements change each
bound by at most an absolute constant factor. We then run the algorithm
with the smaller coefficient; for $q=1$, we use the tree paths. The
comparison uses integer arithmetic and no information about the request
sequence.

For example, if $k\ge n/(\log(n+3))^{\alpha}$ for a fixed $\alpha>0$, then
the competitive ratio of the parity-cover algorithm, apart from the
factor $\rho/\varepsilon$, is at most
$O_\alpha(\log(n+1)[1+\log\log(n+3)])$. If $k=\Omega(n)$, the competitive
ratio of the direct-cut algorithm is $O(\varepsilon^{-1}\log(n+1))$.
These improvements for particular ranges of $k$ do not, however, reduce
the worst-case exponent of the logarithm uniformly over all $k$.

\subsection{The nonuniform algorithm}

Finally, we take
\begin{equation}
 a_*=\ceil{\frac2\varepsilon}\ceil{\log_2(4m)}.
 \label{eq:nonuniform-constants}
\end{equation}
Then $a_*\varepsilon\ge2\ln(4m)$, which is the condition used in
\eqref{eq:nonuniform-repairs}, and $a_*=O(\varepsilon^{-1}\log(m+1))$.
Feasibility gives $k\le m$, and therefore $\theta_*=O(\log(m+1))$.
Substituting these bounds in \eqref{eq:nonuniform-exact}, together with
$\rho=O(\log(n+1))$ from~\cite{Hajiaghayi26}, proves
\eqref{eq:nonuniform-main}. Notice that this choice uses only $m$ and
$\varepsilon$, and not $k$.

\section{The Remaining Exact-Connectivity Question}
\label{sec:exact}

Our results isolate the role of the slack. A deficient separator of size
$\ell$ has $k-\ell$ certified witnesses in every full-$k$ solution, and
our analysis converts these witnesses into a discount on the cost. At
exact connectivity, however, a final separator may have size
$\ell=k-1$, and then this guarantee supplies only one witness. Our proof
loses its discount in this case, while the family of possible separators
can still be large. Hence, no general exact $o(k)\polylog n$ bound
follows from our analysis.

On the other hand, we do not claim that a polynomial dependence on $k$
is necessary for exact connectivity. The logarithmic lower bound that is
known at small fixed connectivity~\cite{GKR12} does not imply such a
dependence. In particular, a
construction with competitive ratio $\Omega(h)$ on $n=\Theta(k2^h)$
vertices remains a lower bound of logarithmic size, even after writing
$h=k^\alpha$. We make no universal lower-bound claim that is polynomial
in $k$.

Nevertheless, there is a concrete obstruction to a natural completion
argument. In \cref{app:independent}, we give a simple graph with one
request, $n=\Theta(k\log k)$, and equal fractional and integral
augmentation optima. Independently rounding one optimal fractional
solution of this instance needs amplification $\Omega(k)$ to succeed with
constant probability, and a smaller amplification may leave a constant
fraction of the optimum to be paid in repairs. This rules out that
particular estimate based on independent rounding. A successful
algorithm for exact completion could instead exploit correlations,
adaptivity, or a different fractional representation.

\section*{AI Use Statement}
This work resulted from extensive interaction between the authors and OpenAI
Astra, with the authors providing suggestions and critically checking Astra's
responses through iterative exchanges. OpenAI Astra was also used in writing
and revising the manuscript through detailed prompts from the authors, with
Gemini Argon assisting in some parts. Anthropic's Claude (Claude Opus 5.5)
was then used to revise the writing of the manuscript, improving its
exposition and making its notation consistent, without changing any of its
results or arguments. The authors directed the process and evaluated the
proposed ideas and arguments.
\ifx\leanrepository\empty
A GitHub link for Lean files formalizing the main theorem will be added
when the formalization is available: [Link].
\else
Lean files formalizing the main theorem are available on GitHub at
\url{\leanrepository}.
\fi

\bibliographystyle{unsrtnat}
\bibliography{draft}
\appendix
\section{A Spanning-Tree Distribution from Average Stretch}
\label{app:trees}

In this appendix, we describe a polynomial-time sampler that relies on
the older published stretch bound, as an alternative to the sharper
logarithmic theorem of~\cite{Hajiaghayi26} that we use in our main
results. The appendix also explains why a statement about average
stretch suffices for the per-edge expectation in \eqref{eq:stretch}. The
reduction uses the spanning-tree game of~\cite{AKPW95} and the
average-stretch algorithm of~\cite{AN19}.

\paragraph{An oracle for weighted average stretch.}
First assume that the graph is simple and that its edge lengths are
positive rationals. We write $s_T(e)=d_T(e)/c_e$ for the stretch of an
edge $e$ in a tree $T$, and we let
$\bar\rho=O(\log(n+1)\log\log(n+3))$ be an integer upper bound on the
average-stretch guarantee of~\cite{AN19} on graphs with $O(n^2)$
vertices. Given a rational distribution $p$ on the $m$ edges, we replace
each edge $e$ by $\mu_e=1+\floor{mp_e}$ parallel copies of the same
length. The total number of copies $\mu(E)$ is at most $2m$, and
$\mu_e\ge mp_e$. Therefore, a tree of small average stretch on this
replicated graph satisfies
\begin{equation}
 \sum_e p_e s_T(e)\le\frac1m\sum_e \mu_es_T(e)\le2\bar\rho.
 \label{eq:oracle}
\end{equation}

The replicated graph has parallel edges, whereas the average-stretch
algorithm is stated for simple graphs. To fix this, we subdivide each
copy into two edges of half the length. The subdivision has
$n+\mu(E)=O(n^2)$ vertices. In a spanning tree of the subdivision, each
new vertex is either a leaf or has both of its incident edges in the
tree. The copies with both halves in the tree project to a spanning tree
on the original vertices. For a copy that projects to a tree edge, the
stretches of the two halves sum to $2$; for any other copy, they sum to
$2+2s_T(e)$. In both cases, the sum is at least $2s_T(e)$. Hence the
average-stretch guarantee on the subdivision implies the inequality used
in \eqref{eq:oracle}.

\paragraph{From the oracle to an explicit distribution.}
We convert this oracle into a small explicit distribution by
multiplicative weights. We start with $\omega_1(e)=1$ for every edge. In
iteration $t$, we form
\[
 p_t(e)=\frac{\omega_t(e)}{\omega_t(E)},\qquad
 p'_t(e)=\frac12p_t(e)+\frac1{2m},
\]
and we obtain a response tree $T_t$ from \eqref{eq:oracle} for the
distribution $p'_t$. Since $p'_t\ge p_t/2$ and $p'_t\ge1/(2m)$, this tree
satisfies
\[
 \sum_ep_t(e)s_{T_t}(e)\le4\bar\rho
 \qquad\text{and}\qquad s_{T_t}(e)\le4m\bar\rho\quad\text{for all }e.
\]
We set $h_t(e)=s_{T_t}(e)/(4m\bar\rho)\in[0,1]$ and update
$\omega_{t+1}(e)=\omega_t(e)(1+h_t(e)/2)$. After $\ell$ iterations, the
total weight and the weight of each edge satisfy
\[
 \ln \omega_{\ell+1}(E)\le\ln m+\frac{\ell}{2m}
 \qquad\text{and}\qquad
 \frac38\sum_{t=1}^{\ell}h_t(e)\le\ln \omega_{\ell+1}(e)\le\ln \omega_{\ell+1}(E).
\]
The second inequality uses $\ln(1+\xi/2)\ge3\xi/8$ for
$\xi\in[0,1]$. We choose the number of iterations $\ell$ to be a power of
two with
\[
 2m\ceil{\log_2(m+1)}\le \ell<4m\ceil{\log_2(m+1)}.
\]
Since $\ell\ge2m\ln m$, the uniform distribution over the $\ell$ response
trees satisfies
\[
 \max_e\frac1{\ell}\sum_{t=1}^{\ell} s_{T_t}(e)\le\frac{32}{3}\bar\rho,
\]
which is \eqref{eq:stretch} with $\rho=32\bar\rho/3$. The construction makes
$O(m\log(m+1))$ calls to the polynomial-time average-stretch algorithm
and uses polynomial-time rational arithmetic. A tree is sampled with
$\log_2 \ell$ fair random bits.

\paragraph{Multigraphs and zero costs.}
For a multigraph, we retain one cheapest edge for each pair of endpoints,
only for the purpose of tree selection. The bound on the expected
distance of this edge also applies to every other edge in its bundle,
since their costs are not smaller. All original edges remain in the
connectivity instance, unless the separate benchmark-preserving reduction
of \cref{lem:pruning} is explicitly used. Zero costs are treated as in
\cref{sec:parity}.

\section{A Limitation of Independent Rounding for Exact Completion}
\label{app:independent}

The slack in our connectivity guarantee is essential to our rounding
analysis. In this appendix, we give an example that explains why exact
completion requires more than independent sampling of a feasible
fractional augmentation. We emphasize that the example concerns one
rounding rule applied to a fixed fractional solution, rather than a lower
bound against online algorithms: there is only one request, and its
minimum-cost integral augmentation is easy to find.

\begin{proposition}
\label{prop:independent-completion}
For integers $w\ge3$ and $\ell\ge2$, there is a simple undirected graph
with
\[
 n=w(2\ell-1)+2,\qquad m=3w\ell,\qquad k=w+1,
\]
and a free base of $s$--$t$ connectivity $w$, such that the optimum of
the capacitated cut LP and the integral augmentation optimum are both
equal to $\ell$. Moreover, there is an optimal fractional solution $y$
with the following property. If the paid edges are selected
independently, each edge $e$ with probability at most $a y_e$, where
$0\le a\le w$, then
\begin{equation}
 \Prb[\lambda(s,t)\ge k]\le w(a/w)^{\ell}.
 \label{eq:independent-success}
\end{equation}
In particular, for $\ell=\lceil\ln w\rceil$, a success probability of at
least a fixed $p_0\in(0,1]$ requires $a\ge(p_0/e)w=\Omega(k)$. In
addition, if $a\le w/4$ and $\ell\ge48\ln w$, then with probability at
least $1-w^{-3}$, the minimum additional cost needed to attain
connectivity $k$ is at least $\ell/2$. Both conclusions apply to families
with $n=\Theta(k\log k)$.
\end{proposition}

\begin{figure}[tb]
\centering
\begin{tikzpicture}[x=1cm,y=1cm,
  vertex/.style={circle,fill=black,inner sep=1.6pt},
  free/.style={draw=black!55,line width=.7pt},
  paid/.style={draw=black,line width=.9pt,dashed},
  every node/.style={font=\small}]
  \foreach \i/\x in {0/0,1/3,2/6,3/9}{
    \node[vertex] (v\i) at (\x,0) {};
  }
  \node[below=5pt] at (v0) {$s$};
  \node[below=5pt] at (v1) {$v_1$};
  \node[below=5pt] at (v2) {$v_2$};
  \node[below=5pt] at (v3) {$t$};
  \foreach \i/\a/\b/\x in {1/0/1/1.5,2/1/2/4.5,3/2/3/7.5}{
    \node[vertex] (z\i) at (\x,1.1) {};
    \node[above=5pt] at (z\i) {$z_{\i}$};
    \draw[free] (v\a)--(v\b);
    \draw[free] (v\a)--(z\i);
    \draw[paid] (z\i)--(v\b);
    \node at (\x,-.24) {$0$};
    \node at (\x-.95,.77) {$0$};
    \node at (\x+.95,.77) {$1$};
  }
\end{tikzpicture}
\caption{A three-stage channel. Solid edges are free; each dashed edge
costs one. The construction has $w$ channels, each with $\ell$ stages,
sharing only $s$ and $t$. A channel contributes its second unit of
connectivity only after every paid edge on that channel is purchased.}
\label{fig:independent-channel}
\end{figure}

\begin{proof}
We take $w$ internally vertex-disjoint $s$--$t$ paths, which we call
\emph{channels}, each consisting of $\ell$ free edges. For every path
edge $uv$, we introduce a private vertex $z_{uv}$ and the detour
$u z_{uv}v$, where the edge $u z_{uv}$ is free and the edge $z_{uv}v$
costs one; see \cref{fig:independent-channel}. All free edges are bought
at the beginning. The graph is simple, and counting the interior vertices
of the channels and the detour vertices gives the stated size.

\paragraph{The fractional and integral optima.}
We set $y_e=1$ on free edges and $y_e=1/w$ on paid edges. Every $s$--$t$
cut crosses an edge of every channel. The corresponding detour also
crosses this cut and contributes a capacity of at least $1/w$,
regardless of the side that contains its private vertex. Thus every such
cut has $y$-capacity at least $w+1$, and the fractional cost is
$w\ell/w=\ell$.

For the reverse inequality, consider the following $\ell$ layer cuts.
The $h$th cut places every path vertex at depths $0,\ldots,h-1$ on the
side of $s$ and every path vertex at depths $h,\ldots,\ell$ on the side
of $t$, and it places each detour vertex on the same side as the path
vertex that precedes it. The boundary of this cut contains exactly $w$
free edges and the $w$ paid edges of stage $h$. Since each free-edge
variable is at most one, every feasible solution of the capacitated cut
LP assigns a total paid capacity of at least one to each layer. The
layers use disjoint sets of paid edges, and therefore their constraints
imply $\LP\ge\ell$. On the other hand, completing all detours on one
channel costs $\ell$ and supplies one additional unit of connectivity.
Hence both optima are equal to $\ell$.

\paragraph{Why independent sampling fails.}
Each stage of a channel has capacity one until its paid detour edge is
bought, and capacity two afterwards. Since the stages of a channel are in
series, the channel has connectivity two exactly when all its paid edges
are bought, and connectivity one otherwise. Different channels share only
their endpoints. Consequently, the total connectivity is exactly $w$
plus the number of completed channels. By independence, the probability
of completing a particular channel is at most $(a/w)^{\ell}$, and a union
bound proves \eqref{eq:independent-success}. If $\ell=\lceil\ln w\rceil$
and the success probability is at least $p_0\in(0,1]$, then
\[
 a\ge w(p_0/w)^{1/\ell}\ge(p_0/e)w,
\]
where we use $\ell\ge\ln w$ and $p_0^{1/\ell}\ge p_0$.

\paragraph{The cost of repairing the sample.}
Let $\xi_j$ be the number of selected paid edges on channel $j$. Since
attaining connectivity $w+1$ requires completing at least one channel,
the minimum repair cost is exactly $\ell-\max_j \xi_j$. For $a\le w/4$,
each $\xi_j$ is stochastically dominated by $\Bin(\ell,1/4)$. A Chernoff
bound and a union bound give
\[
 \Prb\left[\max_j \xi_j\ge \ell/2\right]\le w e^{-\ell/12}\le w^{-3}
 \qquad(\ell\ge48\ln w).
\]
On the complementary event, every channel needs at least $\ell/2$
additional paid edges, which proves the claim about repairs. Taking
$\ell=\lceil\ln w\rceil$ for the first conclusion, or
$\ell=\lceil48\ln w\rceil$ for the second, gives $n=\Theta(k\log k)$.
\end{proof}

In particular, for $a=\sqrt w$ and $\ell=\lceil48\ln w\rceil$, the
expected repair cost is at least $(1-w^{-3})\LP/2$ for all sufficiently
large $w$. Thus no universal bound of the form
$\E[\text{repair cost}]\le \alpha_1 e^{-\alpha_2a}\LP$, with fixed constants
$\alpha_1,\alpha_2>0$, holds for this rounding procedure. The same example applies
to the independent sampling of fundamental cycles when the spanning tree
lies in the free base: selecting a paid fundamental cycle adds exactly its
single paid edge. The example leaves open methods that correlate
purchases, adapt to failed cuts, or choose a different fractional
augmentation.

\end{document}